\documentclass[prx, a4paper, 10pt, twocolumn, floatfix, groupedaddress, aps, longbibliography,nofootinbib, accepted=2021-02-15]{quantumarticle}
\pdfoutput=1
\usepackage{scrextend}
\usepackage{times, mathrsfs, amsmath, amsfonts, graphics, graphicx, cancel, color, amsthm, bbm, mathtools, amssymb, physics}
\usepackage{dsfont}
\usepackage[nice]{nicefrac}
\usepackage[english]{babel}
\usepackage[margin=.75in]{geometry}
\usepackage[T1]{fontenc} 
\usepackage{capt-of}

\usepackage{xfrac,relsize,float}

\usepackage{appendix}

\usepackage[unicode=true, breaklinks=false, pdfborder={0 0 1}, backref=false, colorlinks=true, linkcolor=blue, citecolor=blue]{hyperref}

\def\bra#1{\mathinner{\langle{#1}|}}
\def\ket#1{\mathinner{|{#1}\rangle}}
\def\braket#1{\mathinner{\langle{#1}\rangle}}

\def\BraVert{\egroup\,\mid\,\bgroup}

\definecolor{Blue}{rgb}{0,0,1}
\definecolor{Red}{rgb}{1,0,0}
\definecolor{Green}{rgb}{0,1,0}
\definecolor{darkgreen}{rgb}{0,.7,0}
\definecolor{Purp}{rgb}{.2,0,.2}
\definecolor{white}{rgb}{1,1,1}

\newcommand{\n}{{\nu}}

\newtheorem*{definition}{Definition}
\renewcommand{\thesection}{\Roman{section}} 
\renewcommand{\thesubsection}{\arabic{subsection}}
\usepackage[latin1]{inputenc}
\usepackage{fancyhdr}

\begin{document}
\title{Interference as an information-theoretic game}
\author{Sebastian Horvat}
\email{sebastian.horvat@univie.ac.at}
\affiliation{University of Vienna, Faculty of Physics, Vienna Center for Quantum Science and Technology, Boltzmanngasse 5, 1090 Vienna, Austria}
\author{Borivoje Daki\'{c}}
\email{borivoje.dakic@univie.ac.at}
\affiliation{University of Vienna, Faculty of Physics, Vienna Center for Quantum Science and Technology, Boltzmanngasse 5, 1090 Vienna, Austria}
\affiliation{Institute for Quantum Optics and Quantum Information (IQOQI), Austrian Academy of
Sciences, Vienna, Austria}
\date{17 February 2021}

\begin{abstract}
The double slit experiment provides a clear demarcation between classical and quantum theory, while multi-slit experiments demarcate quantum and higher-order interference theories. In this work we show that these experiments pertain to a broader class of processes, which can be formulated as information-processing tasks, providing a clear cut between classical, quantum and higher-order theories. The tasks involve two parties and communication between them with the goal of winning certain parity games. We show that the order of interference is in one-to-one correspondence with the parity order of these games. Furthermore, we prove the order of interference to be additive under composition of systems both in classical and quantum theory. The latter result can be used as a (semi)device-independent witness of the number of particles in the quantum setting. Finally, we extend our game formulation within the generalized probabilistic framework and prove that tomographic locality implies the additivity of the order of interference under composition. These results shed light on the operational meaning of the order of interference and can be important for the identification of the information-theoretic principles behind second-order interference in quantum theory.   

\end{abstract}

\maketitle

\section{Introduction}\label{Introduction}
As Richard Feynman famously put it, ``the double slit experiment is absolutely impossible to explain in any classical way and has in it the heart of quantum mechanics. In reality, it contains the only mystery'' \cite{feynman}. Indeed, the most common way of introducing quantum theory and its basic tenets is via the double-slit experiment, in which a single particle is shown to produce an interference pattern when sent through two slits, in contrast to classical theory in which this effect is missing.  Much later, Sorkin \cite{sorkin} analyzed multi-slit experiments (i.e. generalizations of the double slit experiment to three or more slits) and noticed that quantum mechanics exhibits second-order interference only, meaning that any measurement pattern produced by a quantum system is reducible to the combination of double-slit interferences. The latter served as a motivation for introducing higher-order interference theories which are defined with respect to the order of interference produced in generalized multi-slit experiments \cite{cubes}. Such theories are usually formulated within the framework of $\textit{generalized probabilistic theories}$ (GPT-s) \cite{single system, higher extension}, with quantum theory being only a particular example. This further motivated the search for a set of intuitive physical principles which could explain why experiments should exhibit at most second-order interference \cite{ruling out, ududec, niestegge, kickback}.\\
In this work we study higher-order interference from the information-theoretic perspective. Firstly, we reconsider the definitions of multi-slit experiments and Sorkin's classification and notice that: (a) the framework is defined only for single particles (or single systems), (b) the interference order is defined with respect to simple operations of blocking/opening the slits, and (c) the (final) measurement refers exclusively to intensity measurements (average number of particles at a particular location on the screen). We proceed by generalizing this setting to an arbitrary number of particles (systems), arbitrary set of (local) operations and arbitrary final measurements.
The framework for higher-order interference is formulated as an information-theoretic game where we analyze generic scenarios involving classical and quantum systems. Our generalization involves many particles (systems) and their ability to win certain parity games. The order of these games and their winning probabilities are directly related to the order of interference (in Sorkin's classification) and to the number of systems involved. This generalization offers a shift in the perspective: higher order interference theories are not defined by which phenomena are allowed (e.g. by the structure of interference patterns), but by which tasks can or cannot be accomplished within the theory (i.e. their information-processing capacity). Furthermore, our study provides a direct relation between the order of interference and number of systems used in the protocol, and can thus provide a (semi) device-independent way of witnessing the number of particles (systems) present in the process.\\
In the final section we provide the information-theoretic formulation within the GPT formalism and we study how the order of interference behaves under the composition of systems. We construct a lower bound on the interference order which is additive under composition. In classical and quantum theory, this lower bound coincides with the upper bound, which in turn shows an interesting connection between the order of interference and the composition law. Finally, motivated by the latter, we prove that in generic local-tomographic theories \cite{bare} the order of interference is indeed additive (under certain additional assumptions that we deem reasonable).\\
Together with an information-theoretic perspective, our findings can be seen in light of paving an alternative way towards understanding the physical principles behind the order of interference of quantum theory: we suspect that an important clue might be provided by the composition of systems and by the tensor-product structure. The latter would thus contribute to the operational and physical understanding of quantum theory \cite{hardy, dakic, masanes, chiri} and provide indications for potential developments of post-quantum theories. Moreover, our work deepens the connection between interference and information processing which has already been alluded to in various contexts involving two-way communication \cite{2way, 2wayexp, infinite}, information speed \cite{infospeed}, quantum acausal processes \cite{Brukner}, superposition of orders \cite{orders} and directions \cite{directions}, quantum combs \cite{combs}, quantum switch \cite {switch} and quantum causal models \cite{causal models}.\\  

\section{Information-theoretic formulation: Parity games}\label{Reformulation}
\begin{figure}
\includegraphics[width=\columnwidth]{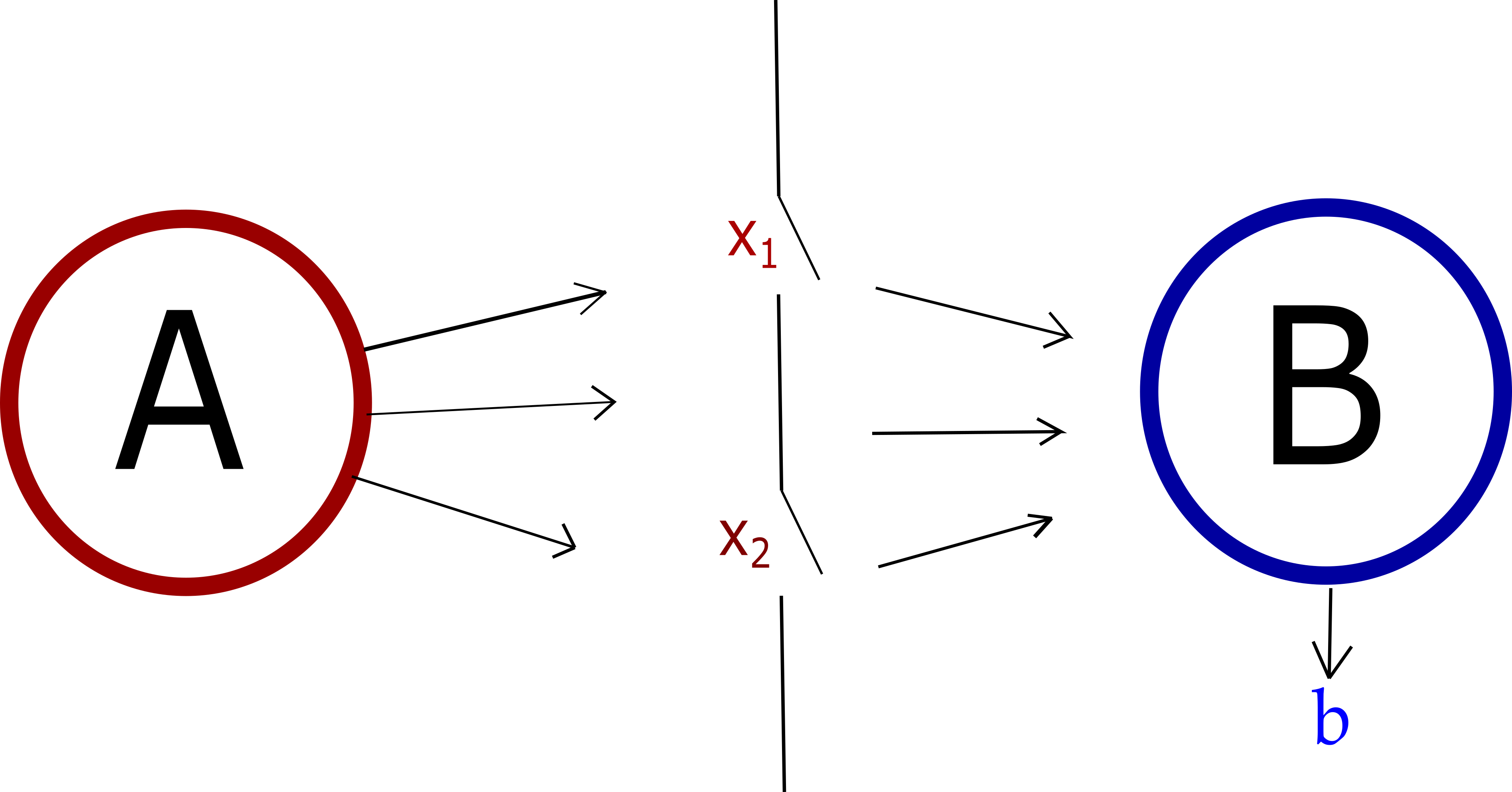}
\caption{Alice and Bob are located at opposite sides of a pierced plate with two slits, each of which can be either open or blocked depending on their input bits $\left\{x_1,x_2\right\}$. Alice sends her particle through the slits towards Bob, who, upon receiving the particle, generates an output bit $b$.}
\label{fig:Fig1}
\end{figure}

\begin{figure}
\includegraphics[width=\columnwidth]{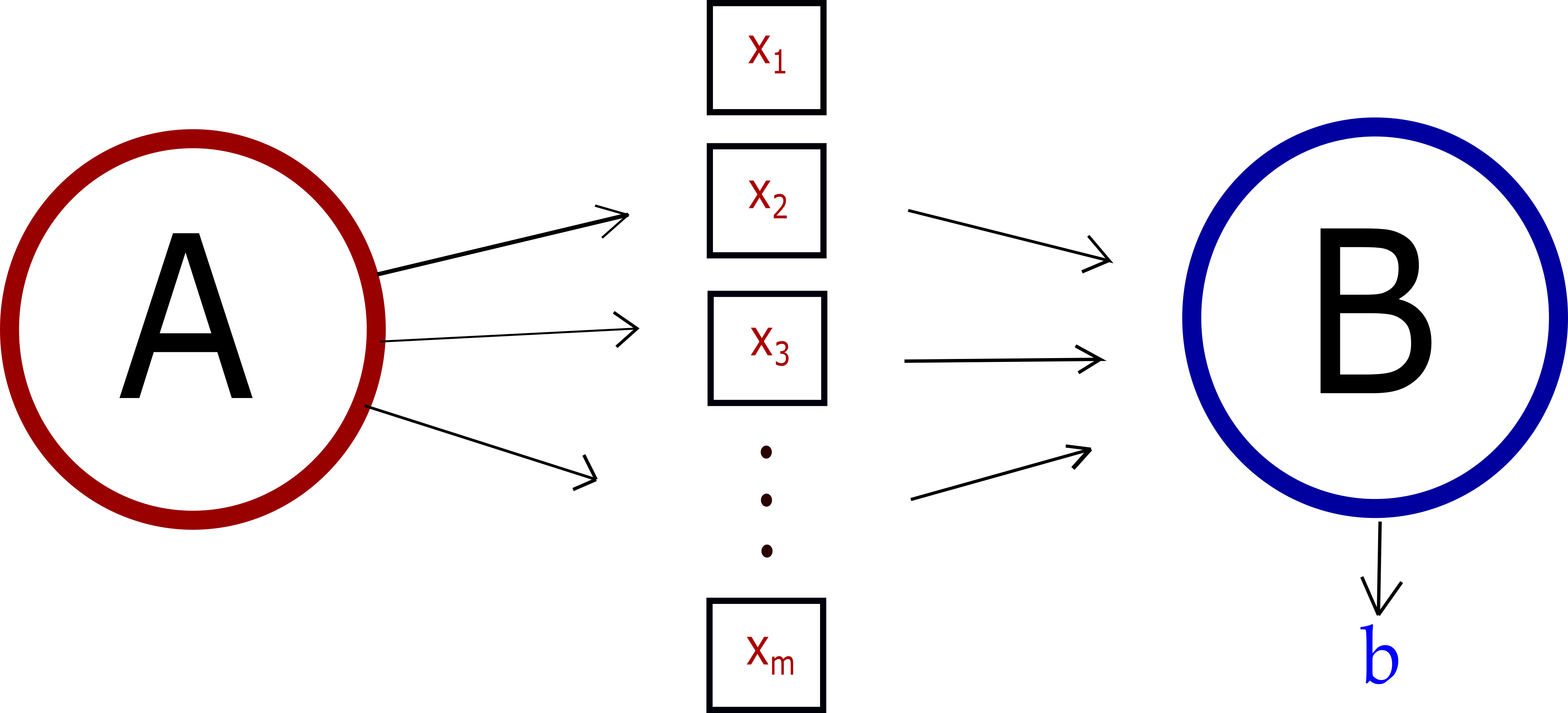}
\caption{ Alice possesses a finite amount of resources that she sends towards Bob through $m$ black boxes. The latter implement arbitrary local operations depending on their pertaining inputs $\left\{x_1,...,x_m\right\}$. Upon receiving the resources, Bob performs an arbitrary operation and generates an output bit $b$.}
\label{fig:Fig2}
\end{figure}

In the standard double-slit experiment a particle is sent on a plate pierced by two parallel slits. After passing through the plate, the particle can be detected on a screen. Each slit can be either open (which we denote with 0), or blocked (which we denote with 1). The figure of merit is the interference term 
\begin{equation}\label{Interf}
I_2=P_{00}-P_{10}-P_{01}+P_{11},
\end{equation}
where $P_{x_1x_2}$ denotes the probability of detecting the particle at a point on the screen given that the slits are in states $x_1$ and $x_2$. Notice that we explicitly included the term which corresponds to the situation in which both slits are closed, despite it being necessarily zero (i.e. $P_{11}=0$). Classical mechanics predicts $I_2=0$, while quantum theory allows the particle to be in spatial superposition, thereby enabling the possibility of generating a non-vanishing $I_2$.\\
In order to reformulate this experiment as an information-theoretic task, let us consider the scenario involving two parties, Alice and Bob, located at opposite sides of the plate, as shown in Figure \ref{fig:Fig1}. 
Let us suppose that the slits are in a certain state (i.e. each slit is either open or blocked) which is completely unknown both to Alice and to Bob; we can thus imagine that an external party (e.g. a referee) has control of the slits and decides whether to block them or not. Moreover, Alice possesses a single particle that she can send towards Bob. On the other side, Bob receives (or not) the particle, performs an arbitrary measurement and outputs a bit $b\in \left\{0,1\right\}$. We introduce the redefined second order interference term as follows
\begin{equation}\label{int2}
\begin{split}
\tilde{I}_2&=\frac{1}{4}\left[P(0|11)-P(0|10)-P(0|01)+P(0|00)\right]=\\
&=\frac{1}{4}\sum_{x_1,x_2=0}^{1} P(b=x_1\oplus x_2|x_1x_2)-\frac{1}{2},
\end{split}
\end{equation}
where $P(b|x_1x_2)$ is the probability that Bob outputs $b$ given that the two slits are in states $x_1$ and $x_2$. Notice that the term corresponding to both slits being closed can now be non-zero, since Bob's measurement is arbitrary, i.e. does not necessarily refer to measuring the probability of the particle inflicting on a point of the screen (intensity measurement). The probability $P(b|x_1x_2)$ is thus a generalization of $P_{x_1x_2}$ from \eqref{Interf} and reduces to the latter only in the case in which Bob performs the intensity measurement. The redefined interference term $\tilde{I}_2$ measures the probability of successfully accomplishing the following task (\textit{parity game}):\\ 
(a) in each run, the slits are set in some state $\left\{x_1,x_2\right\}$ unknown both to Alice and to Bob;\\
(b) Alice prepares her particle and sends it through the slits towards Bob; \\
(c) Bob receives (or not) the particle, and, in order to complete the task, must output the parity $s_{\vec{x}}\equiv x_1\oplus x_2$.\\ 
Bob does not have any prior information about the inputs, which is why a uniform average is taken. We stress that the state of the particle sent by Alice \textit{does not} depend on the inputs $x_1$ and $x_2$, since the latter are unknown to her.\\
Notice that expression \eqref{int2} is formulated in a device-independent way \cite{device}, relating only inputs and outputs, without any mention of their underlying physical realization. It is therefore natural to generalize the scenario by replacing slit operations with generic black boxes, which implement arbitrary local operations depending on their inputs. The boxes can thus perform any operation (e.g. in quantum theory, general completely-positive maps): the only restriction stems from the operations being local. Moreover, instead of constraining Alice to sending only single particles (as it is also the case for standard interference experiments), we can generalize her resources to an arbitrary number of systems and analyze how the winning probability depends on the number of systems used in the process (as we will see in the next section, the game can be won even in classical theory, by using two particles). Additionally, the particles/systems sent by Alice can have any internal structure (e.g. spin) that can be potentially accessed by the black boxes.\\
Analogously to the generalization of the standard double-slit experiment to multi-slit experiments, we can further generalize the scenario to an arbitrary number of boxes $m$. In this case, Alice sends her resources, which consist of $k$ systems, towards Bob, whose task is to output the overall parity of the boxes' inputs, as depicted in Figure \ref{fig:Fig2}. The figure of merit is thus 
\begin{equation}\label{binary definition}
\tilde{I}_m(k)=\frac{1}{2^m}\sum_{x_1,...,x_m=0}^{1} P(b=s_{\vec{x}}|x_1...x_m)-\frac{1}{2},
\end{equation} 
where $\left\{x_1,...,x_m\right\}$ are input bits encoded in the $m$ boxes and $s_{\vec{x}}\equiv \oplus_{i=1}^{m}x_i$ is the overall parity. In the case in which Alice is constrained to sending only single particles, the boxes are implemented as slits and Bob's final operation consists of intensity measurements, the generalized higher-order interference term $\tilde{I}_m(k)$ reduces to the standard higher-order term defined in \cite{sorkin} (up to normalization):
\begin{equation}
I_m= \sum_{x_1,...,x_m=0}^{1} (-1)^{\sum_j x_j} P_{\vec{x}},  
\end{equation} 
where $x_j=0(1)$ corresponds to $j$-th slit being open(closed) and $P_{\vec{x}}$ is the probability of detecting the particle on the screen given that the slits are in state $\vec{x}$.\\
Juxtaposed to the standard definition of $m$-th order interference theories involving the structure of interference patterns produced by single particles in $m$-slit experiments, our information-theoretic extension provides a definition in terms of the probability of winning the parity game involving $m$ boxes by using a finite amount of resources. In addition, the latter formulation enables an analysis of the relation of the order of interference and the number of systems involved in the process (e.g., as we will see in the next section, the $m$-th order parity game can be won using $m$ classical particles). We therefore introduce the characterization of higher-order theories in terms of functions $n(k)$, where $n$ refers to the maximum order of interference that can be exhibited using $k$ systems, i.e. $\tilde{I}_{m=n(k)}(k)\neq 0$ and $\tilde{I}_{m>n(k)}(k)=0$.\\
To recapitulate, the modified multi-slit experiment consists of the following generalizations: (a) instead of one particle, Alice can possess arbitrarily many particles/systems of arbitrary internal structure, (b) the slits are replaced by black boxes which implement generic local operations depending on their pertaining input bits, (c) the screen is replaced by Bob who is supposed to generate an output corresponding to the overall parity of the inputs encoded by the boxes.\\
Finally, we can generalize the input bits $\left\{x_1...,x_m\right\}$ to $dits$, i.e. elements of a set with cardinality $d$. Throughout this manuscript we will assume that $d$ is a prime number, the reason of which will be clearer later. In this case we introduce a class of games in which Bob is supposed to decode one of the $\textit{weighted sums modulo d}$ of the inputs, i.e. $s_{\vec{x},\vec{\n}}^{(d)}\equiv (\sum_{i=1}^{m}\n_i x_i)\text{mod}$ $d$, where $\vec{\n}$ is a dit-string with components $\n_i \in \left\{1...,d-1\right\}$. Given that Alice is using $k$ systems, the generalized interference terms are then 
\begin{equation}\label{Interf mod}
\tilde{I}_{m,\vec{\n},f}^{(d)}(k)=\frac{1}{d^m}\sum_{x_1,...,x_m=0}^{d-1} P\left(b=f(s_{\vec{\n},\vec{x}}^{(d)})|x_1...x_m\right)-\frac{1}{d},
\end{equation}
which are defined for all \textit{reversible} functions $f$ that map dits into dits. Operationally, these functions take into account the potential relabelling of Bob's outputs. Notice that for $d=2$ there was no need for specifying this, since relabelling Bob's output in Eq. \eqref{binary definition} introduces only a minus sign. The games defined in \eqref{Interf mod} offer a definition of higher order interference theories for arbitrary prime $d$. 
\theoremstyle{definition}\label{Def}
\begin{definition}
We define $n(k)$-theories as theories that satisfy the following two properties:\\ 
(i) all processes produce $\tilde{I}_{m>n(k),\vec{\n},f}^{(d)}(k)=0$ for all reversible functions $f$, all dit-strings $\vec{\n}$ with components $\n_i \in \left\{1...,d-1\right\}$, and for all prime $d\geq 2$;\\
(ii) for some prime $d\geq 2$, there exists at least one process that produces $\tilde{I}_{m=n(k),\vec{\n},f}^{(d)}(k)\neq 0$, for at least one dit-string $\vec{\n}$ with components $\n_i \in \left\{1...,d-1\right\}$.
\end{definition} 
Notice that the (non)existence of the required process for a specific outputs' labelling fixed by $f$ implies the (non)existence of the analogous process for any other labelling $f'$ (since Bob can always relabel his outcomes independently of the process). Therefore, throughout this manuscript, we will only construct proofs of existence of processes which exhibit $\tilde{I}_{m=n(k),\vec{\n},f}^{(d)}(k)\neq 0$ for $f$ fixed to be the identity map (i.e. no final relabelling) and we will thereby omit the index $f$. Moreover, whenever we omit the index $\vec{\n}$, we refer to the unit dit-string, i.e. $\n_i=1, \forall i$. Therefore, the term $\tilde{I}_{m}^{(d)}(k)$ will correspond to the specific game in which Bob's goal is to output $s_{\vec{x}}=(\sum_i x_i)\text{mod}$ $d$. We acknowledge that the additional games specified by the labels $\vec{\n}$ and $f$ may at the moment seem redundant; however, we will need to take them into account in the proofs regarding higher-order interference and we will thus keep them in the definition of $n(k)$-theories. As we will show later, all conclusions regarding the order of interference of classical, quantum and higher-order theories will hold for arbitrary prime $d \geq 2$.\\
Winning the parity games can be seen as a truly ``global'' task, in the sense that Bob needs to produce an output that depends on all the local inputs in $\textit{each run}$. For suppose that Bob receives all but one input, say the $j$-th one: in this case, the overall modulo $s_{\vec{x}}=(\sum_i x_i)\text{mod}$ $d$ is completely unknown to Bob. The same conclusion holds if Bob does not receive more inputs or even if he does not receive any information at all. Therefore, the interference term $\tilde{I}_{m,\vec{\n},f}^{(d)}(k)$ is zero irrespectively of whether Bob lacks knowledge about one or about all inputs. Here we can see the importance of assuming that $d$ is a prime number: if it was not, then, in some cases, Bob would be able to guess the overall modulo with a probability greater than $\frac{1}{d}$ even with having no information about one or more inputs. A simple example is the following one: assume that $d=4$, $m=2$, and that Bob is supposed to output the value $(x_1+2x_2)\text{mod}$ $d$ (i.e. $\n_1=1, \n_2=2$); by knowing the value of $x_1$, Bob's winning probability will be $\frac{1}{2}$, which is larger than $\frac{1}{d}$, essentially due to $d$ being divisible by $\n_2$. This would cause several complications in our definitions and proofs, which is why we will simply restrict $d$ to be prime (while still letting it to be arbitrarily large).\\ 
In the subsequent sections, we are going to focus on classical and quantum theory and investigate their power to win the parity games. We will show that the order of interference $n(k)$ satisfies the relation
\begin{equation}
n(k)=kn(1), 
\end{equation}
where $n(1)$ is the order of single-system interference (for classical theory, $n(1)=1$, while for quantum theory, $n(1)=2$) and $n(k)$ is the interference order achievable using $k$ systems. Moreover, we will show that this relation holds in generic GPT-s that satisfy the local tomography principle and certain additional assumptions.\\
Before proceeding further, we will first show an important mathematical property that relates the order of interference to the algebraic order of the probability distributions $P(b|\vec{x})$.

\section{Algebraic order of the probability distributions}\label{math}
In this section we are going to show that any distribution which does not exhibit higher than $n$-th order interference can be written as a linear combination of functions of at most $n$ different inputs, or, in other words, that the algebraic order of $P(b|\vec{x})$ is at most $n$. Here we will consider the $d=2$ case, while the general case is addressed in Appendix \ref{app:A}.\\
Let us consider the parity game involving $m$ boxes and assume that Alice uses one system of order $n<m$. By definition, this means that the interference terms \eqref{binary definition} vanish for all parity games involving more than $n$ boxes, which can be written in a concise form as follows:
\begin{equation}\label{constraints}
\begin{split}
&\tilde{I}_{m,\vec{\sigma}}^{(2)}=\frac{1}{2^m}\sum_{\vec{x}}(-1)^{\sum_j \sigma_j x_j}P(0|\vec{x})=0,\\
&\forall \sigma_j\in\left\{0,1 \right\}, \quad \text{s.t.} \quad \sum_j \sigma_j>n.
\end{split}
\end{equation}
For compactness, we introduced the $m$-component bit string $\vec{\sigma}$, that specifies which interference term the latter equation refers to: if $\sigma_j=1$ for $j=i_1,...,i_l$, the equation states that the order of interference involving boxes $i_1,...,i_l$ is equal to zero ($l$ can be any integer between $n+1$ and $m$).\\
Let us regard $P(0|\vec{x})$ as one component of a vector $\vec{P}$ in a $2^m$-dimensional vector space formed by the tensor product of $m$ two-dimensional spaces, i.e. $P(0|\vec{x})=\vec{e}_{x_1}\otimes...\otimes\vec{e}_{x_m} \cdot \vec{P}$, where $\vec{e}_{x_j=0,1}$ span the $j$-th two-dimensional space. Equations \eqref{constraints} then imply that for all $\sum_j \sigma_j>n$:
\begin{equation}\label{lambda}
\begin{split}
\tilde{I}_{m,\vec{\sigma}}^{(2)}(k=1)&=\frac{1}{2^m}\sum_{\vec{x}}(-1)^{\sum_j \sigma_j x_j}\vec{e}_{x_1}\otimes...\otimes\vec{e}_{x_m} \cdot \vec{P}=\\
&=\frac{1}{2^{m/2}}\vec{f}_{\sigma_1}\otimes...\otimes\vec{f}_{\sigma_m} \cdot \vec{P}=\frac{1}{2^{m/2}}\lambda_{\vec{\sigma}}=0, 
\end{split}
\end{equation}
where we introduced the rotated vectors
\begin{equation}
\vec{f}_{\sigma_j}\equiv \frac{1}{\sqrt{2}}\left(\vec{e}_0+(-1)^{\sigma_j}\vec{e}_1 \right).
\end{equation}  
The terms $\lambda_{\vec{\sigma}}$ are components of vector $\vec{P}$ in the new basis spanned by the rotated vectors $\vec{f}_{\sigma_j}$. Equation \eqref{lambda} then states that the components $\lambda_{\vec{\sigma}}$ are zero if $\sum_j \sigma_j>n$. The probabilities $P(0|\vec{x})$ can thus be expressed as
\begin{equation}
\begin{split}
P(0|\vec{x})&=\left[\bigotimes_{i=1}^{m}\vec{e}_{x_i}\right] \cdot \left[\sum_{\vec{\sigma}} \lambda_{\vec{\sigma}}\bigotimes_{j=1}^{m}\vec{f}_{\sigma_j}\right]=\\
&=\frac{1}{2^{m/2}}\sum_{\substack{\sigma_1,...,\sigma_m \\ \sum_j\sigma_j\leq n}}(-1)^{\sum_j \sigma_j x_j}\lambda_{\vec{\sigma}},
\end{split}
\end{equation}
where we used $\vec{e}_{x_i}\cdot \vec{f}_{\sigma_i}=2^{-1/2}(-1)^{\sigma_ix_i}$. Therefore, $P(0|\vec{x})$ is a linear combination of functions of at most $n$ different inputs:
\begin{equation}
P(0|\vec{x})=\sum_{l=1}^{n}\sum_{i_1,...,i_l} c^{(l)}_{i_1,...,i_l}g^{(l)}(x_{i_1},...,x_{i_l}),
\end{equation}
where we introduced the functions 
\begin{equation}\label{f math}
g^{(l)}(x_{i_1},...,x_{i_l})=(-1)^{\sum_{a=1}^{l} x_{i_a}},
\end{equation}
and the coefficients 
\begin{equation}
c^{(l)}_{i_1,...,i_l}=\frac{1}{2^{m/2}} \vec{\lambda}_{\vec{\sigma}^{(l)}_{(i_1,...,i_l)}}, \quad \left(\vec{\sigma}^{(l)}_{(i_1,...,i_l)}\right)_s=\sum_{r=1}^{l} \delta_{s,i_r}.
\end{equation}
Next, in order to tackle the problem for arbitrary $d$, for each set of generalized interference terms $\left\{\tilde{I}^{(d)}_{m,\vec{\n},f}, \forall \vec{\n}, f\right\}$ we introduce its \textit{dual} terms $\left\{\tilde{J}^{(d)}_{m,\vec{\n},\alpha,b}, \forall \vec{\n}; \forall b \in\left\{0,...,d-1\right\}\right\}$ as 
\begin{equation}
\tilde{J}^{(d)}_{m,\vec{\n},b}=\frac{1}{d^m}\sum_{\vec{x}}(\omega_d)^{ \sum_a  \n_a x_a} P(b|\vec{x}),
\end{equation}
where $\omega_d=e^{i2\pi/d}$ is the $d$-th root of unity. For the $d=2$ case, the dual terms coincide with the ones defined in terms of the game formulation \eqref{binary definition}; however, this is not the case for $d>2$. Nevertheless, in Appendix \ref{app:A} we show that the generalized interference term $\tilde{I}^{(d)}_{m,\vec{\n},f}$ from \eqref{Interf mod} vanishes for all relabellings $f$ and all dit-strings $\vec{\n}$ \textit{if and only if} $\tilde{J}^{(d)}_{m,\n,b}=0$, for all $b=0,...,d-1$ and for all $\vec{\n}$. This allows us to characterize the order of interference by analysing the behaviour of its dual, which can often be more tractable. Indeed, by this method we show in Appendix \ref{app:B} that the conclusion derived in \eqref{f math} holds for arbitrary $d$, i.e. that any distribution that exhibits at most $m$-th order interference can be written as a linear sum of functions which depend on at most $m$ dits.   

\section{Classical resources}\label{classical}
Now we are going to analyze the generalized interference terms achievable within classical theory. Here we will keep the analysis at an informal and intuitive level; a more formal treatment will be given at the end of Section \ref{qm}, where classical theory will be regarded as a special case of quantum theory.\\ 
\textit{Single systems.} For a start, let us focus on the simplest scenario, i.e. the generalization of the double slit experiment, in which Alice sends a single particle through two boxes towards Bob (for now we stick to binary inputs, i.e. $d=2$). Recall that the two boxes can implement arbitrary local operations (not only blocking/opening the slits) labelled by $x_1$ and $x_2$. If the particle is classical, i.e. has a definite trajectory, Bob's output can in each run depend on the state of at most one box. The conditional probabilities describing the process can thus be decomposed as follows:
\begin{equation}
P(b|x_1x_2)=q_1 P_1(b|x_1)+ q_2 P_2(b|x_2),
\end{equation}
where $q_{i}$ is the probability of Alice sending the particle through $i$-th box and $P_{i}(b|x_{i})$ is the conditional probability of Bob outputting $b$ given that the particle was sent through $i$-th box. The modified interference term therefore vanishes:
\begin{equation}
\begin{split}
\tilde{I}_2^{(2)}(k=1)&=\frac{1}{4}\sum_{x_1,x_2=0}^{1}P(b=x_1 \oplus x_2|x_1x_2)-\frac{1}{2}\\
&=\frac{1}{4}\sum_{x_1,x_2=0}^{1} (-1)^{x_1+x_2}P(0|x_1x_2)=0.
\end{split}
\end{equation}
Intuitively, the knowledge about the state of one box does not increase the probability of correctly guessing the inputs' parity with respect to a random guess. The same reasoning holds for arbitrary prime $d$, implying that $\tilde{I}_2^{(d)}(k=1)=0$ for one classical particle.\\ 
\textit{Multiple systems.} Now, what if Alice possesses two classical particles? Let us for the moment assume that the boxes are implemented by slits (as in the original double-slit experiment) and that Alice in each run sends deterministically one particle towards the first slit and the other towards the second slit. If the parity of the slits' states is $x_1 \oplus x_2=0$, then Bob either receives both particles on his side or he receives no particles at all; alternatively, if the parity is $1$, Bob receives exactly one particle. Thus, by simply counting the number of received particles, Bob can in each run determine the inputs' parity and can perfectly accomplish the required task, thereby generating $\tilde{I}_2^{(2)}(k=2)=1/2$. On the other hand, the standard interference definition \eqref{Interf} remains null also for two particles, since the average number of particles received by Bob (or inflicted on the screen) is equal to $1$ regardless of the inputs' parity. This shows the difference between the standard formulation and our game formulation, as the latter allows Bob to measure coincidences, and not only the average particle number (i.e. intensity). \\
We proceed by analysing the fully general scenario with $m$ boxes that implement arbitrary local transformations and assume that Alice's resources consist of $k$ classical objects, be it particles, conglomerates of particles or localized wave packets. For this reason, it is clear that Alice's resources cannot interact with more than $k$ boxes, which means that Bob's output can depend on at most $k$ inputs: if $k<m$, Bob's output is equivalent to a random guess and $\tilde{I}_m^{(d)}(k)=0$, while if $k\geq m$, he can in principle deterministically accomplish the required task and thus generate $\tilde{I}_m^{(d)}(k)=1-1/d$ (if Alice sends one system per box, Bob can in principle retrieve all inputs). The same reasoning holds for all the other interference terms $\tilde{I}_{m,\vec{\n},f}^{(d)}$ (as already noted, this reasoning would not be valid if $d$ was not a prime number).\\
Therefore, classical theory satisfies $n(k)=k$, meaning that $k$ classical systems can exhibit at most $k$-th order interference.

\section{Quantum resources}\label{qm}
Contrasted to classical mechanics, quantum theory allows spatial superpositions of physical objects, which can generate a non-zero interference term $I_2$, even with a single particle. On the other hand, higher order interference terms $I_{j>2}$ defined in multi-slit experiments remain null even in quantum theory (see for instance \cite{sorkin}). In this section we show that analogous statements hold for the generalized interference terms $\tilde{I}_{m,\vec{\n},f}^{(d)}(k=1)$ and provide an extension to more particles (systems). Let us first look into the simplest case, i.e. one particle and two boxes.\\
\subsection{Two boxes}\label{two boxes}
We start by considering the case $d=2$. Let us suppose that Alice possesses a single quantum particle and sends it in spatial superposition towards the two boxes, which implement binary inputs $x_1,x_2$. The quantum state is given by
\begin{equation}
\ket{\psi}_0=\frac{1}{\sqrt{2}} \left( \ket{1}+\ket{2} \right),
\end{equation} 
where $\ket{1}$ and $\ket{2}$ are states corresponding to the two paths. Next, suppose that each box interacts with the particle by applying a local phase-shift $\phi_i=x_i\pi$. After passing through the boxes, the state is
\begin{equation}
\ket{\psi}_{x_1,x_2}=\frac{1}{\sqrt{2}} \left( e^{ix_1\pi}\ket{1}+e^{ix_2\pi}\ket{2} \right).
\end{equation}
Therefore, Bob receives the particle in the following states (up to global phases) 
\begin{equation}
  \ket{\psi}=\begin{cases}
    \frac{1}{\sqrt{2}} \left( \ket{1}+\ket{2} \right), & \text{if $x_1\oplus x_2=0$},\\
    \frac{1}{\sqrt{2}} \left( \ket{1}-\ket{2} \right), & \text{if $x_1\oplus x_2=1$},
  \end{cases}
\end{equation}
which are orthogonal and thus perfectly distinguishable, thereby enabling Bob to deterministically decode the parity $x_1\oplus x_2$ and to produce $\tilde{I}_2^{(2)}(k=1)=1/2$.\\
The previous result holds for binary inputs; now suppose that the parties are playing the modulo game \eqref{Interf mod} specified by dit string $\vec{\n}=(\n_1,\n_2)$, i.e. Bob's goal is to output $s_{\vec{x},\vec{\n}}^{(d)}=(\n_1 x_1+\n_2 x_2) \text{mod } d$ (we assume there is no final relabelling, i.e. that $f$ is the identity; as we already argued, all other cases follow automatically). The players can employ the following strategy.\\
As in the binary case, let Alice prepare her particle in the state 
\begin{equation}
\ket{\psi}_0=\frac{1}{\sqrt{2}} \left( \ket{1}+\ket{2} \right),
\end{equation} 
where $\ket{1}$ and $\ket{2}$ are states corresponding to the two paths, and let the boxes apply local phase shifts as follows
\begin{equation}
\begin{split}
\ket{\psi}_{x_1,x_2}&=\frac{1}{\sqrt{2}} \left( \omega_d^{-\n_1 x_1}\ket{1}+\omega_d^{\n_2 x_2}\ket{2} \right)\\
&=\frac{1}{\sqrt{2}} \left( \ket{1}+\left(\omega_d\right)^{s_{\vec{x},\vec{\n}}^{(d)}}\ket{2} \right),
\end{split}
\end{equation}
where the last equality is valid up to a global phase. Upon receiving the latter state, Bob performs a $d$-outcome POVM $\left\{\Pi_0,...,\Pi_{d-1}\right\}$ corresponding to the $d$ moduli; the interference term is then
\begin{equation}
\begin{split}
\tilde{I}_{2,\vec{\n}}^{(d)}&=\frac{1}{d^2}\sum_{x_1x_2}P\left(b=s_{\vec{x},\vec{\n}}|x_1x_2\right)-\frac{1}{d}\\
&=\frac{1}{d}\sum_s\Tr \left(\Pi_s \rho^{(s)}\right)-\frac{1}{d},
\end{split}
\end{equation}
where
\begin{equation}
\rho^{(s)}\equiv\frac{1}{d}\sum_{\substack{x_1,x_2 \\ s_{\vec{x},\vec{\n}}^{(d)}=s}} \rho_{\vec{x}},
\end{equation}
and $\rho_{\vec{x}}=\ket{\psi}_{x_1,x_2}\bra{\psi}$.\\
Let us choose Bob's POVM to be a ``pretty good measurement'' \cite{pretty good}, which in our case reduces to a set of projectors up to normalization, i.e.
\begin{equation}
\Pi_s=\alpha \rho^{(s)},
\end{equation}
where $\alpha$ is a normalization constant. The latter can be easily calculated by requiring $\sum_s \Pi_s=\mathds{1}$, thereby obtaining $\alpha=\frac{2}{d}$.\\
The interference term is then
\begin{equation}
\tilde{I}_{2,\vec{\n}}^{(d)}= \frac{1}{d}\sum_s\Tr \left( \frac{2}{d} \rho^{(s)} \rho^{(s)}\right)-\frac{1}{d}=\frac{1}{d}>0.
\end{equation}
We have thus shown that a single quantum particle can be used to exhibit second order interference for any $d$. Notice that in this protocol we have not used the internal degree of freedom of the particle and we thus expect that larger interference values may arise in more sophisticated protocols.
\subsection{General case}\label{gneral qm}
Here we analyze the fully general scenario involving arbitrary local operations and arbitrary measurements. We start by constraining Alice's resources to a single quantum system and afterwards we extend our considerations to the case of multiple systems.\\
\subsubsection{Single systems} 
For a start, let us fix the resources to one quantum system, without restricting its internal degrees of freedom. The resource can thus be an electron, an atom or any localized quantum system that can be prepared in coherent spatial superposition using a beam splitter or some more sophisticated mechanism. The local operations implemented by the $m$ boxes will be represented as CP maps acting on the internal degrees of freedom of the system.\\
The most general state Alice can prepare is given by
\begin{equation}\label{input}
\rho_0=\sum_{i,j,k,l} c_{ijkl}\ket{i}\bra{j} \otimes \ket{\phi_k}\bra{\phi_l},
\end{equation}
where $\left\{\ket{i}, i=1,...,m\right\}$ denote states representing defined paths directed towards the $m$ boxes, while $\left\{\ket{\phi_k}\right\}$ span the Hilbert space of the internal degrees of freedom of the system (the dimension of which is arbitrary). The matrix elements $c_{ijkl}$ need to satisfy certain conditions in order for $\rho_0$ to be a legitimate quantum state, but we will not specify them since this constraint is not relevant for what follows.\\
Each box implements a local CP-map that depends on its corresponding input. We denote the Kraus operators representing the action of the $i$-th box with
\begin{equation}
\left\{B_i^{(k)}(x_i), \quad \sum_k B_i^{(k)\dagger}(x_i)B_i^{(k)}(x_i)\leq \mathds{1}\right\},
\end{equation} 
where the operators $B_i^{(k)}(x_i)$ act on the internal Hilbert space of the system. Notice that we allow the transformations to be non-trace-preserving, since they can destroy the system, as for instance in the case of slits. Since the boxes act locally on the system, the total transformation implemented by the boxes is a CP-map with the following Kraus operators:
\begin{equation}\label{cp}
\left\{M^{(k)}_{\vec{x}}=\sum_{i=1}^{m} \ket{i}\bra{i} \otimes B_i^{(k)}(x_i)\right\}.
\end{equation}
The Kraus operators of the total map are thus obtained by conditioning the Kraus operators of the local maps on the path degree of freedom. For example, in the case that the local operators are unitary transformations, for each configuration of inputs there is only one Kraus operator, i.e.
\begin{equation}
M_{\vec{x}}=\sum_{i=1}^{m} \ket{i}\bra{i} \otimes U_i(x_i),
\end{equation}
where $U_i(x_i)$ are unitary operators. Let us now inspect how the general CP-maps in \eqref{cp} act on the input state \eqref{input}. After passing through the boxes, the state of the system is
\begin{equation}\label{rho_general}
\rho_{\vec{x}}=\sum_{k} M^{(k)}_{\vec{x}}\rho_0 M^{(k)\dagger}_{\vec{x}}=\sum_{i,j} \ket{i}\bra{j}\otimes C_{ij}(x_i,x_j),
\end{equation}
where, in order to simplify the expression, we introduced the following operators
\begin{equation}\label{c_ij}
C_{ij}(x_i,x_j)\equiv\sum_{knm}c_{ijnm} B_i^{(k)}(x_i)\ket{\phi_n}\bra{\phi_m}B_j^{(k)\dagger}(x_j).
\end{equation}
Next, let us suppose that Alice and Bob are playing the modulo game \eqref{Interf mod} specified by a dit-string $\vec{\n}$ and with no final relabelling (all other cases follow automatically). Notice that for $m>2$ and for any $\alpha\in \left\{1,...,d-1\right\}$, the following property holds
\begin{equation}\label{eq:c_ij}
\begin{split}
&\quad\sum_{\vec{x}} (\omega_d)^{\alpha\sum_k \n_k x_k} \rho_{\vec{x}}=\\
&= \sum_{i,j} \ket{i}\bra{j} \otimes \sum_{\vec{x}} (\omega_d)^{\alpha\sum_k  \n_k x_k} C_{ij}(x_i,x_j)=0,
\end{split}
\end{equation}
where $\omega_d\equiv e^{i2\pi/d}$. This is so because the tensors $C_{ij}$ depend on at most two inputs and because $\sum_{x_k=0}^{d-1} (\omega_d)^{\alpha \n_k x_k}=0$.\\
Let us define the average taken over states with equal modulo as follows
\begin{equation}
\rho^{(S)}\equiv\frac{1}{d^{m-1}}\sum_{\substack{x_1,...,x_m \\ (\sum_k \n_k x_k)\text{mod d}=S}} \rho_{\vec{x}}.
\end{equation}
As we show in Appendix \ref{app:C}, equations \eqref{eq:c_ij} then imply
\begin{equation}\label{rho}
\rho^{(S)}=\rho^{(S')}\equiv\bar{\rho}, \quad \forall S,S'=0,...,d-1.
\end{equation}
Since all average states $\rho^{(S)}$ are equal, Bob cannot distinguish them by any means. Formally, Bob performs a measurement consisting of $d$ outcomes and represented by a generic POVM $\left\{ \Pi_0,...,\Pi_{d-1}\right\}$. If Alice sends $k=1$ quantum systems, the generalized interference term specified by the dit-string $\vec{\n}$ necessarily vanishes for $m>2$: 
\begin{equation}
\begin{split}
\tilde{I}_{m>2,\vec{\n}}^{(d)}(k=1)&=\frac{1}{d}\Tr [\sum_S \Pi_S \rho^{(S)}] - 1/d \\
&=  \frac{1}{d}\Tr [ \mathbbm{1} \bar{\rho}] -1/d = 0.
\end{split}
\end{equation}
A resource consisting of one quantum system is enough to produce non-vanishing second order interference (for binary inputs, it can even raise the interference term to its maximum possible value); on the other hand, for $m>2$, there is no difference between sending one quantum system and sending no resource at all. This drastic difference can be traced down to \eqref{eq:c_ij}, where the tensors $C_{ij}$ couple at most two inputs $x_i$ and $x_j$. This is because quantum states are described by $(1,1)$ tensors, i.e. density matrices. The latter constraint arises through the Born rule, which essentially sets quantum interference to the second order \cite{cubes}.\\
\subsubsection{Multiple systems and additivity of interference}\label{qm multiple}
Next, let us suppose that Alice's resources consist of more than one quantum system, say, in general, $k$ of them. We assume that the systems are distinguishable and thus described by the tensor product of the single-system Hilbert spaces. The most general state prepared by Alice is then
\begin{equation}
\rho_0=\sum_{\mathbf{i},\mathbf{j},\mathbf{r},\mathbf{l}} c_{\mathbf{i}\mathbf{j}\mathbf{r}\mathbf{l}}\bigotimes_{p=1}^{k}\left[\ket{n^{(p)}_{i_p}}\bra{n^{(p)}_{j_p}}\otimes\ket{\phi^{(p)}_{r_p}}\bra{\phi^{(p)}_{l_p}}\right],
\end{equation}
where $\mathbf{i}$ is short for $(i_1,...,i_k)$ and analogously for the other indices. The vectors $\left\{\ket{n^{(p)}_{i_p}}, \forall i_p=1,...,m\right\}$ span the spatial Hilbert space of the $p$-th system, while $\left\{\ket{\phi^{(p)}_{r_p}}, \forall r_p\right\}$ span its internal degrees of freedom (the dimensionality of which is arbitrary).\\
After passing through the boxes, the state is transformed to
\begin{equation}\label{cij}
\rho_{\vec{x}}=\sum_{\mathbf{i},\mathbf{j}} \bigotimes_{p=1}^{k}\left[\ket{n^{(p)}_{i_p}}\bra{n^{(p)}_{j_p}}\right]\otimes C_{\mathbf{i}\mathbf{j}}(x_{i_1},x_{j_1},...,x_{i_k},x_{j_k}),
\end{equation}
where $\left\{ C_{\mathbf{i}\mathbf{j}} \right\}$ are operators depending on at most $2k$ inputs (defined in the same fashion as in the single-system case, see Eq. \eqref{c_ij}). Here we again assumed that the action of the boxes is represented by local CP-maps conditioned on the path degrees of freedom of the systems (see Appendix \ref{app:D} for details). The crucial difference with respect to the single system case is that the expressions \eqref{eq:c_ij} are now valid only if $m>2k$. Analogously to the single-system case, one comes to the following conclusion
\begin{equation}
\tilde{I}_{m>2k,\vec{\n}}^{(d)}(k)=0.
\end{equation}
We therefore showed that $k$ quantum systems \textit{cannot} produce more than $2k$-th order interference. Now we are going to show that $k$ systems \textit{can} produce $2k$-th order interference. For binary inputs (i.e. $d=2$), Alice and Bob can partition the $2k$ boxes into pairs $\left\{(x_1,x_2),...,(x_{2k-1},x_{2k}) \right\}$ and for each pair use the protocol described in Section \ref{two boxes}; Bob can thus perfectly decode the parity of each pair, which enables him to win the game with unit probability. For $d>2$, the proof is not so straightforward, since the protocol involving one particle and two boxes does not raise the interference term to its maximum value as it does for binary inputs. Nevertheless, in Appendix \ref{app:E} we derive a general result which shows that $k$ generic systems of single-system orders $\left\{n_1,...,n_k\right\}$ \textit{can} produce $(\sum_{i=1}^{k} n_i)$-th order interference, for any $d\geq2$ (this statement is independent of the underlying physical theory). Therefore, $k$ quantum systems (i.e. systems of order two) can produce $2k$-th order interference.\\
According to our classification, the latter results imply that quantum theory satisfies $n(k)=2k$, meaning that $k$ quantum systems can produce at most $2k$-th order interference\footnote{Notice that the same result holds also for indistinguishable quantum systems. In that case, the composite Hilbert space is the (anti)symmetric subspace of the tensor product of the single-system Hilbert spaces. Therefore, the order of interference cannot be higher than the one achievable by indistinguishable systems, since further constraints are imposed on the allowed states and transformations. Moreover, the additivity of the lower bound proved in Appendix \ref{app:E} holds for any systems, including also indistinguishable quantum systems. This implies that all quantum systems satisfy $n(k)=2k$.}.\\
Moreover, since the order of interference depends sharply on the number of particles used in the process, it can be used as a semi-device-independent witness of the number of particles. More precisely, one can extract a lower bound on the number of particles present in the process from purely operational quantities (i.e. conditional probability distributions $P(b|\vec{x})$). This witness is not fully device-independent, since one must still assume that the underlying mechanism obeys quantum theory and that the particle number is well defined (the latter assumption can be violated if the resources consist of e.g. coherent states of light). Additionally, even though the boxes' operations can be mutually spacelike separated (and therefore the boxes cannot mutually signal their local inputs to each other), each box's operation must still be timelike separated from Bob (since the systems must in the end be sent from the boxes towards Bob). Thus, one cannot exclude the possibility that a box signals its local input towards Bob by sending some additional system, which had not been previously sent by Alice. Our particle witness can be seen as similar to device independent dimension witnesses, where instead of the number of particles, one aims to certify the Hilbert space dimension of the underlying system \cite{witness1, witness2}.\\
\textit{Classical theory revisited.} Even though we already showed that classical theory satisfies $n(k)=k$ in Section \ref{classical}, here we want to briefly explicate how the latter can be seen by treating  classical theory as a special case of quantum theory. We simply need to restrict the one-particle-states sent by Alice to be block diagonal in the spatial basis, and the multi-particle-states to be block-diagonal in the configuration space basis. The internal degrees of freedom can remain quantum (since it does not affect our result): here we thus regard ``classical theory'' to mean ``quantum theory with no spatial coherence''. A general classical $k$-particle-state is then:
\begin{equation}
\rho_0=\sum_{\mathbf{i},\mathbf{r},\mathbf{l}} c_{\mathbf{i}\mathbf{r}\mathbf{l}}\bigotimes_{p=1}^{k}\left[\ket{n^{(p)}_{i_p}}\bra{n^{(p)}_{i_p}}\otimes\ket{\phi^{(p)}_{r_p}}\bra{\phi^{(p)}_{l_p}}\right],
\end{equation}
where the notation is the same as before. Analogously to the quantum case, the transformed state is 
\begin{equation}\label{cij}
\rho_{\vec{x}}=\sum_{\mathbf{i}} \bigotimes_{p=1}^{k}\left[\ket{n^{(p)}_{i_p}}\bra{n^{(p)}_{i_p}}\right]\otimes C_{\mathbf{i}}(x_{i_1},...,x_{i_k}),
\end{equation}
where $C_{\mathbf{i}}(x_{i_1},...,x_{i_k})$ are tensors as the ones defined in the quantum case. Since the transformed state is a linear combination of terms, each of which depends on at most $k$ inputs, we recover the result that classical theory (i.e. quantum theory with no spatial coherence) satisfies $n(k)=k$.

\section{Higher-order interference theories}
\subsection{Independent operations}\label{non ent}
In Appendix \ref{app:E} we constructed a proof that $k$ generic systems of orders $\left\{n_1,...,n_k\right\}$ can exhibit $\sum_in_i$-th order interference, for any $d\geq 2$. The latter provides a lower bound which is additive in the interference order. On the other hand, we saw that in classical and quantum theory, this lower bound coincides with the upper bound, i.e. these theories satisfy $n(k)=kn(1)$. In what follows, we will discuss the extendability of this relation to a broader class of theories. In this subsection we will assume that the boxes act independently on each subsystem and show that additivity follows straightforwardly from the principle of local tomography \cite{bare}; in the next subsection we will drop this restriction, but in order to prove our statement we will need to introduce further assumptions.\\
We adopt the standard formalism of \textit{generalized probabilistic theories} (GPTs) \cite{bare}, and suppose that Alice possesses $k$ systems. The state space of the $j$-th system is a convex set that will be denoted with St($A_j$), and we regard it as a subset of a real vector space $A_j$. The set of transformations consists of linear maps acting on $A_j$, and will be denoted with Transf($A_j$). Finally, the set of effects Eff($A_j$) consists of linear functionals that map states into probabilities and can thus be regarded as subset of the dual vector space $A_j^{*}$.\\
Now we proceed with specifying how systems compose. We assume the principle of \textit{local tomography}  \cite{bare}, which states that the state of a composite system can be fully characterized by measurements performed separately on each subsystem. Under this assumption, the composite state space of Alice's systems, which we denote with St($A$), is a subset of the tensor product of the subsystems' vector spaces, i.e. St($A$) $\subset A_1\otimes...\otimes A_k$. Moreover, the composite space of effects Eff($A$) is a subset of the tensor product of the single-systems' dual vector spaces, i.e. $\text{Eff}(A) \subset A_1^{*}\otimes...\otimes A_k^{*}$. Furthermore, as mentioned above, in this subsection we will assume that the boxes act independently on the subsystems; we will discuss the general scenario in the next subsection. We want to show that under the above assumptions, $k$ systems of orders $\left\{n_1,...,n_k\right\}$ cannot exhibit more than $\sum_in_i$-th order interference. Together with the lower bound derived in Appendix \ref{app:E}, this will imply that $k$ systems of orders $\left\{n_1,...,n_k\right\}$ constitute a $\sum_in_i$-th order system. Let us first focus on binary inputs, i.e. $d=2$.\\ 
An arbitrary state $\vec{s} \in$ St($A$) prepared by Alice can be written as a linear combination of the tensor product of single system states: $\vec{s}=\sum_{\mathbf{i}} a_{\mathbf{i}} \vec{s}^{(1)}_{i_1}\otimes...\otimes\vec{s}^{(k)}_{i_k}$, where $\mathbf{i}$ is short for $i_1,...,i_k$, and $a_{\mathbf{i}}$ are real coefficients. The same holds for an arbitrary effect $\vec{f} \in$ Eff($A$) representing Bob's final measurement, i.e. $\vec{f}=\sum_{\mathbf{i}} b_{\mathbf{i}} \vec{f}^{(1)}_{i_1}\otimes...\otimes\vec{f}^{(k)}_{i_k}$. Since we consider only single-system operations, the overall transformation of the boxes on the composite system is given by $T_{\vec{x}}=T_{\vec{x}}^{(1)}\otimes...\otimes T_{\vec{x}}^{(k)}$, where $T_{\vec{x}}^{(j)}\in$ Transf($A_j$) are input-dependent transformations acting on the $j$-th system. For $d=2$, the $m$-th order interference term for an arbitrary process is then:
\begin{equation}\label{int tom}
\begin{split}
&\tilde{I}_{m}^{(2)}(k)=\frac{1}{2^m}\sum_{\vec{x}}(-1)^{\sum_a x_a}P(0|\vec{x})\\
&=\frac{1}{2^m}\sum_{\mathbf{i},\mathbf{j}}a_{\mathbf{i}}b_{\mathbf{j}}\sum_{\vec{x}}(-1)^{\sum_a x_a}\left[\bigotimes_{p=1}^{k}\vec{f}^{(p)}_{j_p}\right] \cdot \left[\bigotimes_{p'=1}^{k}T^{(p')}_{\vec{x}}\vec{s}_{i_{p'}}^{(p')} \right]\\
&=\frac{1}{2^m}\sum_{\mathbf{i},\mathbf{j}}a_{\mathbf{i}}b_{\mathbf{j}}\sum_{\vec{x}}(-1)^{\sum_a x_a} \prod_{p=1}^{k}P^{(p)}_{\vec{x}},
\end{split}
\end{equation}
where $P^{(p)}_{\vec{x}}\equiv\vec{f}^{(p)}_{j_p}\cdot T^{(p)}_{\vec{x}}\vec{s}_{i_{p}}^{(p)}$ are probabilities arising from the single system processes involving $p$-th system (for simplicity we omitted indices $\mathbf{i,j}$). Therefore, we see that due to local tomography and the restriction to single-system operations, the interference term decouples into a linear combination of products of single-system processes. As it was shown in Section \ref{math}, the probability distribution pertaining to any process involving a system of order $n_p$ can be written as a linear combination of functions of at most $n_p$ inputs:
\begin{equation}
P^{(p)}_{\vec{x}}=\sum_{i_1,...,i_{n_p}} c^{(p)}_{i_1,...,i_{n_p}}f^{(p)}(x_{i_1},...,x_{i_{n_p}}).
\end{equation}
Therefore, if $N\equiv\sum_p n_p<m$, the interference term necessarily vanishes:
\begin{equation}
\tilde{I}_{m}^{(2)}(k)=\sum_{i_1,...,i_N}a_{i_1,...,i_N} \sum_{\vec{x}} (-1)^{\sum_a x_a}g(x_{i_1},...,x_{i_N})=0,
\end{equation}
where we introduced for simplicity the coefficients $a_{i_1,...,i_N}$ and functions $g(x_{i_1},...,x_{i_N})$, the exact forms of which are of no relevance.\\
In order to show that the latter results holds for arbitrary prime $d$, we just need to take a look at the dual interference terms defined in Section \ref{math}. Given our assumptions, one arrives to the generalization of \eqref{int tom}:
\begin{equation}
\tilde{J}_{m,\vec{\n},b}^{(d)}(k)=\frac{1}{d^m}\sum_{\mathbf{i},\mathbf{j}}a_{\mathbf{i}}b_{\mathbf{j}}\sum_{\vec{x}}(\omega_d)^{\sum_a \n_a x_a} \prod_{p=1}^{k}P^{(p)}_{b,\vec{x}}.
\end{equation}
By assumption, the probabilities $P^{(p)}_{b,\vec{x}}=\vec{f}^{(p)}_{b,j_p}\cdot T^{(p)}_{\vec{x}}\vec{s}_{i_{p}}^{(p)}$  pertaining to $p$-th system do not exhibit more than $n_p$-th order interference and can thus be expressed as a linear combination of functions depending on at most $n_p$ inputs, as we showed in Appendix \ref{app:B}. Therefore, if $N\equiv\sum_p n_p<m$, then $\tilde{J}_{m,\vec{\n},b}^{(d)}(k)$ vanishes for all $\vec{\n}, b$. Since the dual formulation is equivalent to the game formulation (see Appendix \ref{app:A}), this implies that the $k$ systems cannot be used to achieve more than $\sum_j n_j$-th order interference for any prime $d$.\\
On a side note, one may wonder whether the above proof could be applied to general non-independent transformations, since local tomography implies that all transformations can be written as a linear combination of independent transformations, i.e. $T_{\vec{x}}=\sum_{\mathbf{i}}c_{\mathbf{i}}(\vec{x})\bigotimes_{p=1}^{k}T^{(p)}_{i_p\vec{x}}$. However, due to the coefficients $c_{\mathbf{i}}(\vec{x})$ being explicitly dependent on $\vec{x}$, our proof would not be valid in general. Therefore, in order to drop the assumption of independent transformations, we will characterize the locality of the boxes' operations in the next subsection.

\subsection{Non-independent operations and locality}\label{ent}
In the previous section we proved that under the assumptions of (i) local tomography, and (ii) independent operations, a system composed of $k$ systems of orders $(n_1,...,n_k)$ is a $(\sum_j n_j)$-th order system. Notice that within the proof we did not need to specify anything about the transformations $T_{\vec{x}}$ and we did not impose the notion of locality of the boxes' actions. On the other hand, in quantum (and classical) theory we proved a stronger statement: additivity follows without assumption (ii); however, in that case we explicitly implemented the notion of locality by conditioning the boxes' actions on the path degrees of freedom of the systems. In what follows, we will provide an analogous analysis in the context of GPTs, thereby dropping assumption (ii). We will start by defining the relevant concepts for single systems.\\
In order to generalize the quantum proof from Section \ref{qm} to GPTs, we need to introduce controlled operations, which means that we will treat the degrees of freedom of the systems as composed of the ``path'' degree of freedom (which will serve as the control) and the ``internal'' degrees of freedom (on which the information will be encoded). Let us start by considering a single-system $A$. We denote the state space of the control degrees of freedom with St($A^{(c)}$) $\subset A^{(c)}$, and the state space of the internal degrees of freedom with St($A^{(int)}$)  $\subset A^{(int)}$, where $A^{(c)}$ and $A^{(int)}$ are real vector spaces. Assuming local tomography, the total state space of system $A$ will then be St($A$) $\subset A^{(c)}\otimes A^{(int)}$. Analogously, we can define the spaces of effects as subspaces of the dual vector spaces, i.e. $\text{Eff}(A^{(c)}) \subset  A^{(c)*}$, $\text{Eff}(A^{(int)}) \subset A^{(int)*}$, and $\text{Eff}(A) \subset A^{(c)*}\otimes A^{(int)*}$. Now we will characterize the state space of the control degree of freedom.\\
\\
\textbf{Assumption 0.} Let us assume that there exists an $m$-outcome measurement for the control degree of freedom corresponding to the following set of effects $\left\{ \vec{f}^{(1)},...,\vec{f}^{(m)}\in \text{Eff}(A^{(c)});\quad \sum_i \vec{f}^{(i)}=\vec{u}^{(c)}\right\}$, where $\vec{u}^{(c)}$ is the unit effect, i.e. the effect that satisfies $\vec{u}^{(c)}\cdot \vec{v}=1,\forall \vec{v}\in$ $\text{St}(A^{(c)})$. The physical meaning of the latter is that each effect $\vec{f}^{(i)}$ gives the probability of the system being found at box $i$. We will thus refer to this measurement as the ``which-box measurement''. \\
\\
Let us also define subsets of boxes $J= \left\{j_1,,...,j_{|J|}\right\} \subseteq I$, where $I \equiv \left\{1,2,...,m\right\}$, and introduce the coarse-grained effects $\vec{f}^{(J)}\equiv \sum_{i=1}^{|J|} \vec{f}^{(j_i)}$, with $|J|$ being the cardinality of $J$. Effect $\vec{f}^{(J)}$ thus outputs the probability of the system being found at the subset of boxes $J\subseteq I$. We will say that an arbitrary state $\vec{v} \in \text{St}(A^{(c)})$ is \textit{localized at subset $J$} if:
\begin{itemize}
\item $\vec{f}^{(J)}\cdot \vec{v}=1$\footnote{Or equivalently: $\forall J',  J' \cap J = \emptyset$   $ \rightarrow $   $\vec{f}^{(J')}\cdot \vec{v}=0$.};
\item $\forall J', \quad J' \subset J$   $ \rightarrow $   $\vec{f}^{(J')}\cdot \vec{v}>0$.
\end{itemize}
The first point of this definition simply say that ``a system being localized at some particular subset'' means that ``the probability of finding the system outside of this subset is null''. The second point is needed in order to select the minimum possible subset which satisfies the first point and thus select a unique subset.\\ 
We will now consider a basis $S=\left\{\vec{s}_1,...,\vec{s}_{r^{(c)}} \right\}$ of $A^{(c)}$, with $\vec{s}_i \in$ St($A^{(c)}$), and $r^{(c)}$ being the dimension of $A^{(c)}$. If the theory describing our system was classical, then $r^{(c)}=m$, and we could choose a basis consisting of states localized at each box, i.e. $\vec{f}^{(i)}\cdot \vec{s}_j=\delta_{ij}$. On the other hand, if we consider quantum theory, then $r^{(c)}=m^2$, and we can construct a basis $S=\left\{\vec{s}_{j_1j_2},\forall j_1,j_2=1,...,m \right\}$ such that each of its elements $\vec{s}_{j_1j_2}$ is localized at subset $\left\{j_1,j_2\right\}$. One example of such a basis is the following: states $\vec{s}_{jj}$ correspond to the classical states localized at each box, i.e. $\vec{s}_{jj}=\ket{j}\bra{j}$, whereas, for $j_1<j_2$, states $\vec{s}_{j_1j_2}$ and $\vec{s}_{j_2j_1}$ correspond to the superposition states $\vec{s}_{j_1j_2}=\ket{j_1j_2}_x\bra{j_1j_2}$ and $\vec{s}_{j_2j_1}=\ket{j_1j_2}_y\bra{j_1j_2}$, where $\ket{j_1j_2}_x \sim \left(\ket{j_1}+\ket{j_2}\right)$ and $\ket{j_1j_2}_y \sim \left(\ket{j_1}+i\ket{j_2}\right)$ (see for instance \cite{hardy}). Here we see that the state-space of quantum theory can be spanned by states, each of which is localized at most at two boxes. On the other hand, all states in classical theory can be spanned by states, each of which is localized at one box. Tentatively, one may guess from the latter that there is a strong relation between the ``localizability'' of the state-space of a theory with respect to the ``which-box measurement'' and the order of interference of the theory\footnote{A similar reasoning motivated the construction of the density cube theory which is a third-order-interference theory obtained as a modification of quantum theory, by replacing density matrices (tensors with two indices) with tensors with three indices (the so called density cubes). Translated to our framework, such a state space is spanned by states, each of which is localized at most at three boxes \cite{cubes}.}. In what follows, we will analyze this relation and show that it does indeed hold under certain assumptions. However, we will first formalize the notion of the ``localizability'' with respect to the ``which-box measurement'' by introducing the notion of ``locality order''.\\
\\
\textbf{Locality order.} Let us focus on a generic state-space of the control degree of freedom $\text{St}(A^{(c)})$ (while still assuming that there exists the which-box measurement, i.e. Assumption 0), and consider an arbitrary basis $S=\left\{\vec{s}_1,...,\vec{s}_{r^{(c)}} \right\}$ of $A^{(c)}$, with $\vec{s}_i \in$ St($A^{(c)}$). The elements of $S$ can be localized at various boxes' subsets, and these subsets consist in general of different numbers of boxes (i.e. the subsets have different cardinalities). Let us define $l_S \in \mathbb{N}$ to be the maximum among these numbers, meaning that each element of $S$ is localized within at most $l_S$ different boxes. Formally, for each basis $S=\left\{\vec{s}_1,...,\vec{s}_{r^{(c)}} \right\}$ we define the parameter $l_S$ in the following way:
\begin{equation}
\begin{split}
\forall i, \quad \exists \left( J_i \subseteq I \right), \quad \text{s.t.} \quad \left(\vec{s}_i \text{ is localized at } J_i\right) \& \left(|J_i|\leq l_S \right).\\  
\end{split}
\end{equation}
Naturally, different basis choices $S$ correspond to different values of $l_S$. We will define the \textit{locality order} $L$ as the \textit{minimum} $l_S$ among all possible basis, i.e. $L\equiv \min_S l_S$, and denote a basis that achieves this value of $l_S$ with $S^{*}$ (there can be many different basis that achieve the minimum value; we will simply take $S^{*}$ to be one of them). Therefore, the parameter $l_S$ is associated to each basis $S$, whereas the locality order $L$ is associated to the theory, together with the ``which-box measurement''. Quantum theory satisfies $L=2$ (with respect to the specific which-box measurement as defined in Assumption 0), since there exists a complete basis whose elements consist of superpositions of at most two boxes (see the example from the previous paragraph), and there is no basis with all of its elements localized at single boxes (classical theory of course satisfies $L=1$). Note that a hypothetical theory could also satisfy $L=m$, meaning that any basis must contain at least one ``completely de-localized'' state. Now we are ready to state our main definitions and assumptions, which will formalize the notion of local operations.\\
\\
\textbf{Assumption 1}. Here we will introduce the notion of locality of the boxes' operations. As before, we will label with $S^{*}=\left\{\vec{s}_1,...,\vec{s}_{r^{(c)}}\right\}$ one of the basis of the control degree of freedom, which satisfy $l_{S^{*}}=\min_S l_S=L$. Each of the states $\vec{s}_n$ is localized at subset $J_n=\left\{j_1^{(n)},...,j_{|J_n|}^{(n)}\right\}$, with $|J_n| \leq L$. Let us now suppose that a system $A$ is initially prepared in the product state $\vec{s}_i \otimes \vec{g}$, where $\vec{s}_i \in S^*$, and $\vec{g} \in \text{St}(A^{(int)})$. The initial state is thus localized within the subset of boxes $J_i \subseteq I$. We deem it natural to assume that the boxes act in such a way that (i) the transformed state is still localized within subset $J_i$ (i.e. the boxes cannot ``delocalize'' the system), and (ii) the transformed state cannot depend on inputs pertaining to boxes outside of $J_i$ (i.e. if the probability of finding the system at some box is null, then the system should not be affected by that box). We will formalize point (i) by requiring that the boxes implement transformations $T_{\vec{x}}$ on the input state $(\vec{s}_i \otimes \vec{g})$ in such a way that the output state satisfies
\begin{equation}\label{point1}
(\vec{f}^{(J)} \otimes \vec{e}) \cdot T_{\vec{x}}(\vec{s}_i \otimes \vec{g})=0,
\end{equation}
for all coarse-grained effects $\vec{f}^{(J)}$ with $J \cap J_i = \emptyset$, and arbitrary effects $\vec{e}\in \text{Eff}(A^{(int)})$. Equation \eqref{point1} thus states that one cannot steer the control degree of freedom of the output state into a state which is localized outside of the initial subset $J_i$ (i.e. this is what we mean that the output state is localized within $J_i$). The most general form of the action of the transformations $T_{\vec{x}}$ that satisfies the above requirements is the following:
\begin{equation}\label{transfx}
T_{\vec{x}}(\vec{s}_i \otimes \vec{g})=\sum_{\substack{n \\ J_n \subseteq J_i}} \vec{s}_{n} \otimes T^{(i,n)}_{\vec{x}^{(J_i)}}\vec{g},
\end{equation}
where the sum runs over those basis states pertaining to $S^*$ which are localized within the initial subset, i.e. each basis state $\vec{s}_{n} \in S^*$ is localized within subset $J_n$, with $J_n \subseteq J_i$. The transformations $T^{(i,n)}_{\vec{x}^{(J_i)}}$ are linear operators on $A^{(int)}$ and depend only on inputs contained within subset $J_i$, i.e. $\vec{x}^{(J_i)}\equiv \left(x_{j^{(i)}_1},...,x_{j^{(i)}_{|J_i|}}\right)$, where we used the notation $J_i=\left\{j^{(i)}_1,...,j^{(i)}_{|J_i|}\right\}$. Moreover, all additional coefficients that may appear in front of each term of the sum are for simplicity absorbed in the operators $T^{(i,n)}_{\vec{x}^{(J_i)}}$ (also, note that the vectors $T^{(i,n)}_{\vec{x}^{(J_i)}}\vec{g} \in A^{(int)}$ may generally lie outside of the internal state space $\text{St}(A^{(int)})$).\\ 
\\
\\
\textbf{Relation between the locality order and the order of interference}. Now we will show that Assumption 1 and the principle of local tomography imply a strong relation between the order of interference and the locality order. Consider a general instance of our game which consists of Alice preparing her system in an arbitrary state, sending it through the boxes and Bob performing a generic measurement, thereby producing the conditional probabilities $P(b|\vec{x})$. Let us introduce a basis for the internal degrees of freedom $\left\{\vec{g}_1,...,\vec{g}_{r^{(int)}} \right\}$, where $r^{(int)}$ is the dimension of the internal vector space $A^{(int)}$. According to the principle of local tomography, any initial state $\vec{v} \in  \text{St}(A)$ can be written as
\begin{equation}
\vec{v}=\sum_{i,j}c_{ij}\vec{s}_i \otimes \vec{g}_j,
\end{equation}
where $c_{ij}$ are real coefficients, and $\vec{s}_i \in S^*$. The boxes implement a transformation $T_{\vec{x}}$, and Bob's $d$-outcome measurement is modelled via a generic set of effects $\left\{\vec{E}_0,...,\vec{E}_{d-1}\in \text{Eff(A)}, \sum_i \vec{E}_i=\vec{u} \right\}$, where $\vec{u}$ is the unit effect on the total state-space. The conditional probability is thus
\begin{equation}\label{L=k}
\begin{split}
P(b|\vec{x})&=\sum_{i,j} c_{ij} \vec{E}_b \cdot \left( T_{\vec{x}} \vec{s}_{i} \otimes \vec{g}_j\right)=\\
&=\sum_{\substack{i,j,n \\ J_n \subseteq J_i}} c_{ij} \vec{E}_b \cdot \left(\vec{s}_{n} \otimes T^{(i,n)}_{\vec{x}^{(J_i)}}\vec{g}_j \right),
\end{split}
\end{equation}
where in the second equality we used Assumption 1. Each transformation $T^{(i,n)}_{\vec{x}^{(J_i)}}$ depends only on inputs within subset $J_i$. Therefore, for a given theory with locality order $L$, the probabilities $P(b|\vec{x})$ are at most of algebraic order $L$. Now we can see a direct connection to the order of interference: indeed, as proven in Section \ref{math} and Appendix \ref{app:B}, if a distribution exhibits $k$-th order interference, then its algebraic order is $k$. Thus, if a system \textit{can} exhibit $k$-th order interference, then Assumption 1 implies that the theory satisfies $L\geq k$.\\
Next, we will introduce a further assumption which will relate the possible transformations implementable by the boxes to the locality order of the theory, and show that together with the previous assumptions, it implies $L=k$.\\
\\
\textbf{Assumption 2}. Let us suppose that the locality order of our system is $L$, meaning that all states can be written as linear combinations of states, each of which is localized in subsets of at most $L$ boxes. Moreover, there exist states which cannot be decomposed into linear combinations of states, each of which is localized at \textit{less} than $L$ boxes. We assume the following: there exists at least one state $\vec{v}\in \text{St}(A)$ and one set of transformations $\left\{T_{\vec{x}}, \forall \vec{x}\right\}$, such that the transformed state $T_{\vec{x}}\vec{v}$ is of algebraic order $L$, i.e. it cannot be decomposed into a combination of terms, each of which depends on less than $L$ inputs. Let us heuristically illustrate the meaning and motivation of this assumption with a simple example involving two boxes and a system described by a theory with locality order $L=2$. Let us take for the moment that our assumption \textit{does not} hold, i.e. that all states and transformations in the theory give rise only to states with algebraic order 1. This means that for any state $\vec{v}\in \text{St}(A)$ and any set of transformations $\left\{T_{x_1x_2}, \forall x_1,x_2\right\}$, the following holds:
\begin{equation}\label{ass2}
T_{x_1,x_2}\vec{v}=\sum_i \vec{s}_i \otimes T^{(v,i)}_{x(i)} \vec{g}_i, 
\end{equation}
where $\vec{s}_i \in S^*$, $\vec{g}_i \in \text{St}(A^{(int)})$. The inputs $x(i)$ can be either $x_1$ or $x_2$ depending on $i$, which makes each of the operators $T^{(v,i)}_{x(i)}$ depend only on one input. On the other hand, if we considered a system of locality order $L=1$ (e.g. classical theory), then a general output state $T_{x_1,x_2}\vec{v}$ within that theory would have a similar form to the one in Equation \eqref{ass2}: it would still be a linear combination of terms, each of which depends only on one input. Thus, if our assumption did not hold, a system of locality order 2 would encode the inputs in the qualitatively same way as a system of order 1. In order to introduce a difference between systems of different orders, we will assume that a system of order $L$ can produce a state with algebraic order $L$. This assumption is rather technical and admittedly requires further physical justification; however, in this work we will take it at face value and show that, together with the other assumptions, it implies additivity of interference under composition. Moreover, note that the assumption is satisfied by quantum theory: namely, if the system is prepared in a superposition of two boxes, there exist local transformations that encode the two inputs on the state (e.g. local phase gates) in such a way that the transformed state cannot be decomposed into a linear combination of terms, each of which depends only on one input. On the other hand, superpositions of more than two boxes still give rise only to distributions of algebraic order 2, which is related to the fact that the quantum density matrix is a tensor with two indices (which essentially sets the locality order to $L=2$).\\
\\
$\mathbf{L=k}$. Let us now briefly show how the conjunction of our assumptions implies $L=k$. Namely, Assumption 2 directly implies that if the system \textit{cannot} exhibit more than $k$-th order interference, then $L\leq k$, since for $L> k$ there would exist states and transformations that produce probabilities of algebraic order larger than $k$ which would exhibit more than $k$-th order interference.
On the other hand, as we have seen before, Assumption 1 and the principle of local tomography imply $L\geq k$. Hence, the conjunction of the latter statements implies that the locality order of a $k$-th order system is $L=k$.\\
\\
\textbf{Assumption 3}. Up until now, the discussion has been focused only on single systems; now we will specify how the boxes act on composite systems. In sub-section \ref{non ent} we had assumed that the boxes act independently on the subsystems; here we want to drop this assumption. Let us consider a system composed of two subsystems. We will label the basis that achieve the minimal locality order for the first and second subsystem respectively with $S_1^*$ and $S_2^*$. Suppose that the system is prepared in the following product state $\vec{s}_{i_1}\otimes \vec{s}_{i_2}\otimes \vec{g}_1 \otimes \vec{g}_2 \in A_1^{(c)}\otimes A_2^{(c)} \otimes A_1^{(int)} \otimes A_2^{(int)}$, where $\vec{g}_1$ and $\vec{g}_2$ are arbitrary internal states of the two subsystems, and $\vec{s}_{i_1} \in S_1^*$, $\vec{s}_{i_2}\in S_2^*$, are localized respectively at subsets $J_{i_1}=\left\{j_1^{(i_1)},...,j_{|J_{i_1}|}^{(i_1)}\right\} \subseteq I$ and $J_{i_2}=\left\{j_1^{(i_2)},...,j_{|J_{i_2}|}^{(i_2)}\right\} \subseteq I$. Since the two subsystems are localized respectively at $J_{i_1}$ and $J_{i_2}$, we will say that the composite system is localized at $J_{i_1} \cup J_{i_2}$. Now we will introduce the action of the boxes on the composite system as a direct generalization of Assumption 1. Namely, we will demand that (i) the composite control degree of freedom of the transformed state can be steered only into states localized within $J_{i_1} \cup J_{i_2}$, and (ii) the transformed state depends only on inputs contained within subset $J_{i_1} \cup J_{i_2}$. Analogously to Assumption 1., the first part states that the boxes cannot further ``delocalize'' the system, while the second part states that if the probability of finding either of the subsystems at some box is 0, then the composite system should not be affected by the action of that box. Thus we see that this assumption states that if two subsystems of a composite system are localized at subsets $J_{i_1}$ and $J_{i_2}$, then the transformations acts on the composite system, as if the latter was a single system localized at subset $J_{i_1} \cup J_{i_2}$. The most general form of the transformations $T_{\vec{x}}$ that satisfies the above requirements is
\begin{equation}\label{transf comp}
\begin{split}
&T_{\vec{x}}\left(\vec{s}_{i_1} \otimes \vec{s}_{i_2} \otimes \vec{g}_1 \otimes \vec{g}_2\right)\\
&=\sum_{\substack{n_1n_2 \\ (J_{n_1} \cup J_{n_2})\\ \subseteq (J_{i_1} \cup J_{i_2})}} \vec{s}_{n_1} \otimes \vec{s}_{n_2} \otimes T^{(\mathbf{i}\mathbf{n})}_{\vec{x}^{(J_{i_1},J_{i_2})}}\left(\vec{g}_1 \otimes \vec{g}_2\right),
\end{split}
\end{equation} 
where we introduced the input strings
\begin{equation}\label{strings}
\vec{x}^{(J_{i_1},J_{i_2})}\equiv (x_{j_1^{(i_1)}},...,x_{j_{|J_{i_1}|}^{(i_1)}},x_{j_1^{(i_2)}},...,x_{j_{|J_{i_2}|}^{(i_2)}}),
\end{equation}
and the bold index $\mathbf{n}$ stands for $n_1n_2$. Each of the transformations $T^{(\mathbf{i}\mathbf{n})}_{\vec{x}^{(J_{i_1},J_{i_2})}}$ is a linear operator on $A^{(int)}_1 \otimes A^{(int)}_2$ that depends on inputs $\vec{x}^{(J_{i_1},J_{i_2})}$. The sum in Equation \eqref{transf comp} runs over states which are localized within the union of the initial subsets, i.e. states $\vec{s}_{n_1}$ and $\vec{s}_{n_2}$ are localized respectively at subsets $J_{n_1}$ and $J_{n_2}$, where the latter satisfy $(J_{n_1} \cup J_{n_2}) \subseteq (J_{i_1} \cup J_{i_2})$.
Notice that transformations that entangle the subsystems are not excluded and are to be expected if the initial states of the control systems of the two subsystems are localized within overlapping subsets of the boxes, i.e. if $J_{i_1} \cap J_{i_2} \neq \emptyset$.\\ 
\\
\textbf{Additivity of interference}. Now we will show that the conjunction of Assumption 3 with the previous assumptions implies that interference is additive under composition of systems. Let us consider a system composed of two $k$-th order subsystems. As we have shown before, the principle of local tomography  and Assumptions $1 \& 2$ imply that the locality orders of both subsystems are $L=k$. Consequently, this enables us to choose the following basis for the composite control state space: $\left\{\vec{s}_{i_1}\otimes \vec{s}_{i_2}, \forall i_1,i_2=1,...,r^{(c)}\right\}$, where the single-system states $\vec{s}_{i_1}$ and $\vec{s}_{i_2}$ are localized respectively at subsets $J_{i_1}=\left\{j^{(i_1)}_1,...,j^{(i_1)}_{|J_{i_1}|)}\right\}$ and $J_{i_2}=\left\{j^{(i_2)}_1,...,j^{(i_2)}_{|J_{i_2}|}\right\}$, with $|J_{i_1}|\leq k$ and $|J_{i_2}|\leq k$. From here it is simple to show that two $k$-th order systems cannot exhibit more than $2k$-th order interference. Namely, due to the assumption of local tomography, the most general initial state prepared by Alice is
\begin{equation}
\vec{v}=\sum_{\mathbf{i}\mathbf{k}}c_{\mathbf{i}\mathbf{k}}\vec{s}_{i_1} \otimes \vec{s}_{i_2} \otimes \vec{g}_{k_1} \otimes \vec{g}_{k_2},
\end{equation} 
where $\left\{\vec{g}_{k_1} \otimes \vec{g}_{k_2}, \forall k_1,k_2\right\}$ forms a basis for the internal state space of the composite system. The bold indices $\mathbf{i},\mathbf{k}$ stand for $i_1,i_2,k_1,k_2$, and $c_{\mathbf{i}\mathbf{k}}$ are real coefficients. The composite system is then sent through the boxes, which implement the input-dependent transformation $T_{\vec{x}}$. Finally, Bob performs a measurement $\left\{\vec{E}_0,...,\vec{E}_{d-1}\in \text{Eff(A)}, \sum_i \vec{E}_i=\vec{u} \right\}$, where $\vec{u}$ is the unit effect on the total state-space. The conditional probabilities thus follow:
\begin{equation}
\begin{split}
&P(b|\vec{x})=\sum_{\mathbf{i}\mathbf{k}} c_{\mathbf{i}\mathbf{k}} \vec{E}_b \cdot \left[ T_{\vec{x}} \left(\vec{s}_{i_1} \otimes \vec{s}_{i_2} \otimes \vec{g}_{k_1} \otimes \vec{g}_{k_2}\right) \right]\\
&=\sum_{\mathbf{i}\mathbf{k}\mathbf{n}} c_{\mathbf{i}\mathbf{k}} \vec{E}_b \cdot \left[\vec{s}_{n_1} \otimes \vec{s}_{n_2} \otimes T^{(\mathbf{i}\mathbf{n})}_{\vec{x}^{(J_{i_{1}}J_{i_{2}})}}\left(\vec{g}_{k_1} \otimes \vec{g}_{k_2} \right) \right].
\end{split}
\end{equation}
Since $\vec{x}^{(J_{i_{1}}J_{i_{2}})}$, as defined in Equation \eqref{strings}, is a string of at most $2k$ different inputs, the algebraic order of the conditional probabilities arising from such processes is bounded by $2k$, which implies that two $k$-th order systems produce at most $(2k)$-th order interference. The procedure can be trivially generalized to an arbitrary number of systems, by composing them two-by-two, thus arriving to the conclusion that the upper bound on interference is additive under the composition of systems. Together with the result in Appendix \ref{app:E} which proves the additivity of the lower bound of interference under composition, this completes the proof that a system composed of subsystems of orders $\left\{n_1,...,n_p\right\}$ is a $(\sum_{i=1}^{p} n_i)$-th order system. Note that this result is valid for the inputs $x_i$ being elements of a set of cardinality $d$, with $d$ being an arbitrary prime number.\\
\\
We will now briefly summarize and emphasize the main assumptions and results of this section. In Subsection \ref{non ent} we had proved the additivity of interference under the assumptions of local tomography and independent operations, without referring at all to the locality of the boxes' operations. In Subsection \ref{ent}, in order to drop the second assumption, we implemented the notion of locality via controlled operations, as a natural generalization of the quantum and classical cases. Besides the principle of local tomography and the assumption of the existence of the which-box measurement (Assumption 0), we introduced three further assumptions on single-(Assumptions 1 $\&$ 2) and composite-system transformations (Assumption 3). Finally, notice that in this section and overall in the whole manuscript, we treated locality via controlled operations; one may however adopt a different view and from the very start associate to each space(time) region a system, the transformations on which would automatically be regarded as local (in quantum theory, such a formulation would correspond to second-quantization and to local quantum physics \cite{local qm}, which makes it more compatible with relativity).

\section{Conclusion and Outlook}
In this work we introduced a class of information-theoretic games, which generalize standard multi-slit interference experiments. The order of interference of a theory is then seen as the (im)possibility of accomplishing certain information-processing tasks using finite resources. These games essentially characterize how much information can be decoded from a physical system given that the information was encoded in a global property (parity or modulo of the inputs) of local pieces of information (local inputs). We showed that within quantum theory, the order of interference of $k$ systems is $2k$; it would be interesting to inspect potential connections of this result to superdense coding, which states that $k$ qubits can be used to send at most $2k$ bits. Moreover, the game formulation can provide a (semi)-device-independent witness of the particle number. Finally, even though our work is mainly focused on conceptual issues, it might inspire novel experimental tests of higher-order interference (albeit now redefined in terms of our game); for instance, one could replace slits with boxes implementing generic unitary operations and observe how the order of interference scales with the number of systems. This could then complement the vast experimental search for higher-order interference \cite{exp1, exp2, exp3, exp4}. Another interesting direction is to take the boxes' inputs to be concrete spatio-temporal variables, as in Ref. \cite{spatio-temp}, and see how this constrains the order of interference achievable with a fixed amount of resources (e.g. one may perhaps find connections between the order of interference and the dimensionality of space).\\
So far there have been several attempts at explaining the order of interference of quantum theory; however, we deem that a physically intuitive explanation has not yet been obtained. The reason for this might be that the sole question is leading us in the wrong direction. Instead of asking ourselves why does quantum theory behave in a particular way in multi-slit experiments (or GPT generalizations thereof), we could ask why does quantum theory restrict the amount of globally-encoded information (parity/modulo of locally encoded bits/dits) that can be retrieved from a system. Moreover, what is the relation between the (im)possibility of such a decoding and the number of systems used for encoding the information? Is there any physical argument for why interference should be additive under composition (e.g. violation of the no-signalling principle)? How does one even define the notion of \textit{number of systems} in a device-independent scenario (or ought one to quantify the resources in an alternative manner)? We plan to investigate these and further questions in future works.

\begin{acknowledgements}
The authors thank the three anonymous reviewers for their insightful comments and suggestions which greatly improved both the content and the form of this manuscript. The authors thank Markus P. M{\"u}ller, Joshua Morris and Nicol{\'a}s M. S{\'a}nchez for helpful discussions. B.D. acknowledges support from an ESQ Discovery Grant of the Austrian Academy of Sciences (OAW) and the Austrian Science Fund (FWF) through  BeyondC-F7112. S.H. acknowledges support from an ESQ Discovery Grant of the Austrian Academy of Sciences (OAW).
\end{acknowledgements}

\renewcommand\refname{Bibliography}

\appendix

\onecolumn

\section{Equivalence between the game formulation and its dual}\label{app:A}
In this section we will show that the formulation of higher order interference theories in terms of winning probabilities of the "modulo" games
\begin{equation}\label{game}
\tilde{I}_{m,\vec{\n},f}^{(d)}(k)=\frac{1}{d^m}\sum_{x_1,...,x_m=0}^{d-1} P\left(b=f(s_{\vec{\n},\vec{x}}^{(d)})|x_1...x_m\right)-\frac{1}{d}=0, \quad  \forall \vec{\n}, \forall f
\end{equation}
is equivalent to its dual formulation in terms of the dual interference terms
\begin{equation}\label{dual}
\tilde{J}^{(d)}_{m,\vec{\n},b}=\frac{1}{d^m}\sum_{\vec{x}}(\omega_d)^{\sum_a  \n_a x_a} P(b|\vec{x})=0, \quad \forall \vec{\n}, \quad \forall b\in\left\{0,...,d-1\right\},
\end{equation}
where $\omega_d=e^{i2\pi/d}$ is the $d$-th root of unity.\\
For convenience, let us first introduce the dual terms with an added index $\alpha \in \left\{1,...,d-1\right\}$:
\begin{equation}
\tilde{J}^{(d)}_{m,\vec{\n},b,\alpha}\equiv\frac{1}{d^m}\sum_{\vec{x}}(\omega_d)^{\alpha\sum_a \n_a x_a} P(b|\vec{x}).
\end{equation}
We will show that for any dit-string $\vec{\n}$, the following equivalence holds:
\begin{equation}\label{equiv}
\left\{ \tilde{I}_{m,\vec{\n},f}^{(d)}=0, \quad \forall f	\right\}  \longleftrightarrow \left\{ \tilde{J}^{(d)}_{m,\vec{\n},b,\alpha}=0, \forall b\in\left\{0,...,d-1\right\}, \forall \alpha \in \left\{1,...,d-1\right\}\right\}
\end{equation}
Since the order of interference is defined as the vanishing of the interference terms \eqref{game} for all $\vec{\n}$, \eqref{equiv} would imply that the dual formulation of interference \eqref{dual} is equivalent to the game formulation \eqref{game} (this is so because if the RHS of equivalence \eqref{equiv} holds \textit{for all} $\vec{\n}$, then the dual conditions \eqref{dual} also hold, since the index $\alpha$ becomes redundant). In what follows, we will prove that \eqref{equiv} does hold indeed.\\
In order to simplify the notation, let us introduce the $d \times d$ matrix $P$, whose elements are defined as\footnote{Notice that the matrix $P$ depends on the dit-string $\vec{\n}$, but we do not write it explicitly in the notation in order to simplify the expressions.}
\begin{equation}
P_{bs}=\frac{1}{d^{m-1}}\sum_{\substack{\vec{x} \\ s_{\vec{\n},\vec{x}}^{(d)}=s}}P(b|\vec{x}).
\end{equation}
The normalization of probabilities implies that $P$ is a stochastic matrix, i.e. 
\begin{equation}\label{norm}
\sum_b P_{bs}=1, \quad \forall s.
\end{equation}
The game formulation equations on the left side of the equivalence in \eqref{equiv} can be written as
\begin{equation}
\frac{1}{d^m}\sum_{x_1,...,x_m=0}^{d-1} P\left(b=f(s_{\vec{\n},\vec{x}}^{(d)})|x_1...x_m\right)=\frac{1}{d}, \quad \forall f \quad\rightarrow \quad \sum_s P_{f(s)s}=1, \quad \forall f, 
\end{equation}
which we rewrite succintly as
\begin{equation}\label{game cond}
\Tr [\Pi_f P]=1, \quad \forall f,
\end{equation}
where $\Pi_f \equiv \sum_s \ket{f(s)}\bra{s}$ ranges over all $d$-dimensional permutation matrices (the most general finite dimensional reversible transformations are indeed permutations).\\
On the other hand, the dual conditions on the right side of equivalence \eqref{equiv} assume the following form
\begin{equation}\label{dual cond}
(PF)_{lk}=\delta_{k,0} \frac{1}{\sqrt{d}} \sum_j P_{lj} ,
\end{equation}
where $F$ is the $d$-dimensional Fourier matrix with elements
\begin{equation}
(F)_{lk}=\frac{1}{\sqrt{d}} (\omega_d)^{lk}.
\end{equation}
We will first show that the dual conditions \eqref{dual cond} imply the information-theoretic conditions \eqref{game cond}. Assuming that the the dual conditions \eqref{dual cond} hold, we evaluate the trace
\begin{equation}
\Tr [\Pi_f P]=\Tr [F^{\dagger}\Pi_f P F]= \sum_{kl} (F^{\dagger}\Pi_f)_{kl} (PF)_{lk}=\sum_{l} (F^{\dagger}\Pi_f)_{0l} (PF)_{l0}=1, \forall f.
\end{equation}
The first step follows from the unitarity of $F$ and the ciclicity of the trace, the third step is due to the dual conditions \eqref{dual cond}, and the last step follows from $(F^{\dagger}\Pi_f)_{0l}=1/\sqrt{d}$. Therefore, conditions \eqref{dual cond} imply \eqref{game cond}.\\
In order to prove the converse statement, let us introduce the following new basis $\beta=\left\{\ket{E}, \ket{e_{1}},...,\ket{e_{d-1}}\right\}$, where $\ket{E}$ has the following form
\begin{equation}
\ket{E}=1/\sqrt{d} (1,1,...,1)^T,
\end{equation}
and the remaining vectors span the orthogonal subspace (they can for instance be chosen as the rows/columns of the Fourier matrix). The normalization conditions \eqref{norm} can then be written as
\begin{equation}
\bra{E}P=\bra{E},
\end{equation}
which implies that the matrix $P$ in the new basis has the following form
\begin{equation}
P= \ket{E}\bra{E} + \sum_j q_j \ket{e_j}\bra{E} + \bar{E},  
\end{equation}
where $q_j$ are arbitrary complex numbers and $\bar{E}$ is a matrix that has support only on the subspace orthogonal to $\ket{E}$.\\
In the new basis $\beta$, all permutation matrices have the following form
\begin{equation}
\Pi_f= \ket{E}\bra{E} + \Delta_f,  
\end{equation}
where $\Delta_f$ is again a matrix with no support on $\ket{E}$. The latter follows from the fact that the representation of the permutation group that we are using is reducible to the direct sum of a one-dimensional representation (spanned by the vector $\ket{E}$ which is invariant under all permutations) and a $(d-1)$-dimensional irreducible representation (here given by $\Delta_f$). Assuming that the game-conditions \eqref{game cond} hold, we obtain the following
\begin{equation}\label{burn}
1=\Tr [\Pi_fP]= 1+ \Tr [\Delta_f \bar{E}] \rightarrow \Tr[\Delta_f \bar{E}]=0, \quad \forall f.
\end{equation}
Since the matrices $\Delta_f$ provide an irreducible representation of the permutation group, the Burnside Theorem \cite{burnside} implies that they span the set of all $(d-1)$-dimensional matrices; therefore, Equation \eqref{burn} implies that $\bar{E}$ is necessarily zero. The matrix $P$ is thus
\begin{equation}
P= \ket{E}\bra{E} + \sum_j q_j \ket{e_j}\bra{E} = \ket{q}\bra{E},  
\end{equation}
where we introduced the vector $\ket{q}\equiv \ket{E}+ \sum_j q_j\ket{e_j}$. Therefore, the sought matrix $PF$ is 
\begin{equation}
(PF)_{kl}=\braket{l|q}\bra{E}F \ket{k}= \delta_{k,0} \frac{1}{\sqrt{d}} \sum_j P_{lj},
\end{equation}
where we used $\sum_{k=0}^{d-1} (\omega_d)^k=0$ and $\left\{\ket{k}; k=0,...,d-1\right\}$ are the original basis vectors. Hence, equivalence \eqref{equiv} holds, and thus the game formulation \eqref{game} is equivalent to the dual formulation \eqref{dual}.
Besides being an interesting mathematical result, the exact equivalence between these two formulations will be useful for showing that any distribution that exhibits at most $m$-th order interference can be written as a linear combination of functions with at most $m$ inputs. This will consequently be the necessary ingredient for proving that Local Tomography implies the additivity of the order of interference under the composition of systems.

\section{Algebraic order for arbitrary $d$}\label{app:B}
In this Appendix we are going to extend the mathematical property of Section \ref{math} to arbitrary $d$.\\ 
Let us consider the modulo games involving $m$ boxes and assume that Alice uses a system of order $n<m$. In Appendix \ref{app:A} we showed that this is equivalent to the following constraints
\begin{equation}\label{constraints_d}
\begin{split}
&\tilde{J}^{(d)}_{m,\vec{\n},b}=\frac{1}{d^m}\sum_{\vec{x}}(\omega_d)^{\sum_j \n_j x_j}P(b|\vec{x})=0,\\
&\forall b\in\left\{0,1,...,d-1\right\}, \n_j\in\left\{0,1,...,d-1\right\}, \quad \text{s.t.} \quad \sum_j \delta_{\n_j,0}<(m-n).
\end{split}
\end{equation}
For compactness, we allow the dit string $\vec{\n}$ to range over all dit-strings with less than $(m-n)$ null components. This notation specifies which interference terms the equation refers to: if $\n_j\neq 0$ for $j=i_1,...,i_l$, the equation states that the order of interference involving boxes $i_1,...,i_l$ is equal to zero ($l$ can be any integer between $n+1$ and $m$).\\
Let us regard $P(b|\vec{x})$ as one component of a vector $\vec{P}_b$ in a $d^m$-dimensional vector space formed by the tensor product of $m$ $d$-dimensional spaces, i.e. $P(b|\vec{x})=\bigotimes_{i=1}^{m}\vec{e}_{x_i} \cdot \vec{P}_b$, where $\left\{\vec{e}_{x_j}, x_j=0,...,d-1\right\}$ span the $j$-th $d$-dimensional space. Equations \eqref{constraints_d} then imply:
\begin{equation}\label{lambda_d}
\tilde{J}^{(d)}_{m,\vec{\n},b}=\frac{1}{d^m}\sum_{\vec{x}}(\omega_d)^{\sum_j \n_j x_j}\bigotimes_{i=1}^{m}\vec{e}_{x_i} \cdot \vec{P}_b=\frac{1}{d^{m/2}}\bigotimes_{i=1}^{m}\vec{f}_{\n_i} \cdot \vec{P}_b=\frac{1}{d^{m/2}}\lambda_{\vec{\n},b}=0,
\end{equation}
where the rotated vectors (rows/columns of the $d$-dimensional Fourier matrix) are defined as
\begin{equation}
\vec{f}_{\n_j}\equiv \frac{1}{\sqrt{d}}\sum_{x_j=0}^{d-1} (\omega_d)^{\n_j x_j} \vec{e}_{x_j},
\end{equation}  
and the coefficients $\lambda_{\vec{\n},b}$ are the components of the vector $\vec{P}_b$ in the corresponding basis. Equation \eqref{lambda_d} states that the components $\lambda_{\vec{\n},b}$ vanish for all $\vec{\n}$ with more than $n$ non-zero components. The probabilities $P(b|\vec{x})$ can thus be expressed as
\begin{equation}
P(b|\vec{x})=\left[\bigotimes_{i=1}^{m}\vec{e}_{x_i}\right] \cdot \left[\sum_{\vec{\n}} \lambda_{\vec{\n},b}\bigotimes_{j=1}^{m}\vec{f}_{\n_j}\right]=\frac{1}{d^{m/2}}\sum_{\substack{\n_1,...,\n_m \\ \sum_j\delta_{\n_j,0}\geq (m-n)}}(\omega_d)^{\sum_j \n_j x_j}\lambda_{\vec{\n},b},
\end{equation}
where we used $\vec{e}_{x_i}\cdot \vec{f}_{\n_i}=\frac{1}{\sqrt{d}}(\omega_d)^{\n_ix_i}$. Since the sum runs only over strings $\n$ with at most $n$ non-zero components, $P(b|\vec{x})$ can be written as a function of at most $n$ inputs, for every $b$.\\
Therefore we have proved that the $n$-th order interference dual conditions \eqref{constraints_d}, or equivalently, the information-theoretic formulation \eqref{Interf mod}, provide a set of necessary and sufficient conditions for a distribution to be decomposable into a linear combination of at most $n$ inputs. Hence, the algebraic order of the distribution $P(b|\vec{x})$ is at most $n$.

\section{Proof of Equation \eqref{rho}}\label{app:C}
In this Appendix we will show that equations
\begin{equation}\label{app:c_ij}
\sum_{\vec{x}} (\omega_d)^{\alpha\sum_k \n_k x_k} \rho_{\vec{x}}=0, \quad \forall \alpha \in \left\{1,...,d-1\right\}
\end{equation}
imply the following relations
\begin{equation}\label{app:rho}
\rho^{(S)}=\rho^{(S')}, \quad \forall S,S'=0,...,d-1,
\end{equation}
where $\rho^{(S)}$ is defined as
\begin{equation}
\rho^{(S)}\equiv\frac{1}{d^{m-1}}\sum_{\substack{x_1,...,x_m \\ (\sum_k \n_k x_k)\text{mod d}=S}} \rho_{\vec{x}}.
\end{equation}
For a start, Equations \eqref{app:c_ij} can be written as
\begin{equation}\label{app:c_ij2}
\sum_{S=0}^{d-1} (\omega_d)^{\alpha S} \rho^{(S)}=0, \quad \forall \alpha \in \left\{1,...,d-1\right\}.
\end{equation}
Next, if we introduce the vector $\vec{\rho}=(\rho^{(0)},\rho^{(1)},...,\rho^{(d-1)})^T$, the latter equations can be cast in the following form
\begin{equation}\label{matrix}
F \vec{\rho} = \sqrt{d}\bar{\rho}(1,0,...,0)^T,
\end{equation}
where $F_{kl}=1/\sqrt{d}(\omega_d)^{kl}$ is the $d$-dimensional Fourier matrix and $\bar{\rho}=1/d\sum_S \rho^{(S)}$. Taking the inverse of Equation \ref{matrix} (i.e. applying $F^{\dagger}$ from the left) we obtain
\begin{equation}\label{result}
\vec{\rho} = \bar{\rho}(1,1,...,1)^T,
\end{equation}
which is equivalent to 
\begin{equation}
\rho^{(S)}=\rho^{(S')}=\bar{\rho}, \quad \forall S,S'=0,...,d-1.
\end{equation}

\section{Details of the proof of additivity in quantum theory}\label{app:D}
Here we will write explicitly the expressions for the transformations invoked in the proof of the additivity of interference under composition in quantum theory (Section \ref{qm multiple}). We will denote the ``control'' degree of freedom of $j$-th system with $H^{(c_j)}$ (the control is simply the path degree of freedom), and its ``internal'' degree of freedom with $H^{(int_j)}$. The total Hilbert space $H$ of the $k$ systems is a tensor product of the single-system Hilbert spaces, i.e. 
\begin{equation}
H=H^{(c)} \otimes H^{(int)},
\end{equation}
where $H^{(c)}\equiv \bigotimes_{j=1}^{k} H^{(c_j)}$ is the composite of the control/path degrees of freedom, and $H^{(int)}\equiv \bigotimes_{j=1}^{k} H^{(int_j)}$ is the composite of the internal degrees of freedom.\\
Analogously to the single-system case, we introduce the following set of Kraus operators that act on the composite internal Hilbert space $H^{(int)}$:
\begin{equation}
\left\{B_{\mathbf{i}}^{(s)}(x_{i_1},...,x_{i_k}), \quad \sum_s B_{\mathbf{i}}^{(s)\dagger}(x_{i_1},...,x_{i_k})B_{\mathbf{i}}^{(s)}(x_{i_1},...,x_{i_k})\leq \mathds{1}\right\},
\end{equation}
where $\mathbf{i}$ stands for $i_1,...,i_k$, and $\mathds{1}$ is the identity operator on $H^{(int)}$. We do not impose any further restrictions on the latter operators, thereby including also operations that entangle the internal degrees of freedom of the subsystems. The total transformation implemented by the boxes for a given set of inputs $\vec{x}$ will be represented with the following Kraus operators that act on the total Hilbert space $H$:
\begin{equation}
\left\{M^{(s)}_{\vec{x}}=\sum_{\mathbf{i}} \bigotimes_{p=1}^{k} \ket{n^{(p)}_{i_p}}\bra{n^{(p)}_{i_p}} \otimes B_{\mathbf{i}}^{(s)}(x_{i_1},...,x_{i_k})\right\},
\end{equation}
where $\left\{\ket{n^{(p)}_{i_p}}, \forall i_p=1,...,m \right\}$ forms a basis of the control-space of $p$-th system $H^{(c_p)}$.\\
Let us now suppose that the total system is prepared in the following state
\begin{equation}
\rho_0=\sum_{\mathbf{i},\mathbf{j},\mathbf{r},\mathbf{l}} c_{\mathbf{i}\mathbf{j}\mathbf{r}\mathbf{l}}\bigotimes_{p=1}^{k}\left[\ket{n^{(p)}_{i_p}}\bra{n^{(p)}_{j_p}}\otimes\ket{\phi^{(p)}_{r_p}}\bra{\phi^{(p)}_{l_p}}\right],
\end{equation}
where $\mathbf{i}$ is short for $i_1,...,i_k$ (and analogously for the other indices), and vectors $\left\{\ket{\phi^{(p)}_{r_p}}, \forall r_p\right\}$ span the internal space of $p$-th system $H^{(int_p)}$.\\
The boxes then act on the system as follows:
\begin{equation}
\rho_{\vec{x}}=\sum_{s} M^{(s)}_{\vec{x}}\rho_0 M^{(s)\dagger}_{\vec{x}}=\sum_{\mathbf{i},\mathbf{j}} \bigotimes_{p=1}^{k}\left[\ket{n^{(p)}_{i_p}}\bra{n^{(p)}_{j_p}}\right]\otimes C_{\mathbf{i}\mathbf{j}}(x_{i_1},x_{j_1},...,x_{i_k},x_{j_k}),
\end{equation}
where 
\begin{equation}
C_{\mathbf{i}\mathbf{j}}(x_{i_1},x_{j_1},...,x_{i_k},x_{j_k})=\sum_{s,\mathbf{r},\mathbf{l}} c_{\mathbf{i}\mathbf{j}\mathbf{r}\mathbf{l}} B_{\mathbf{i}}^{(s)}(x_{i_1},...,x_{i_k}) \left( \bigotimes_{p=1}^{k} \ket{\phi^{(p)}_{r_p}}\bra{\phi^{(p)}_{l_p}} \right) B_{\mathbf{j}}^{(s)\dagger}(x_{j_1},...,x_{j_k}).
\end{equation}
As stated in Section \ref{qm multiple}, this implies that $k$ quantum systems constitute a $2k$-th order system, since the probability distributions $P(b|\vec{x})$ arising from processes involving $k$ systems will necessarily be at most of algebraic order $2k$.
\section{A lower bound on interference of generic composite systems}\label{app:E}
In this Appendix we are going to show that two systems, labelled by $(A,B)$ and respectively of single-system orders $(n_A,n_B)$, can be used to achieve $(n_A+n_B)$-th order interference.\\
By definition, there exist two processes involving single-systems $A$ and $B$ that can win the modulo games with $n_A$ and $n_B$ boxes with some probabilities $q_A>1/d$ and $q_B>1/d$. For concreteness, suppose that Alice and Bob play the modulo game for the unit dit-string $\n_i=1, \forall i$, with $(n_A+n_B)$ boxes, i.e. Bob is supposed to output $s_{\vec{x}}^{(d)}=(\sum_j x_j) \text{mod}$ $d$. The proof for any other $\vec{\n}$ is completely analogous. The protocol proceeds as follows. Alice sends system $A$ to the first $n_A$ boxes (i.e. containing inputs $\left\{x_1,...,x_{n_A}\right\}$) and system $B$ through the other $n_B$ boxes (which encode inputs $\left\{x_{n_A+1},...,x_{n_A+n_B}\right\}$). Upon receiving the systems, Bob performs two separate measurements with outcomes $b^{(A)}$ and $b^{(B)}$ and produces a final output $b=\left(b^{(A)}+b^{(B)}\right) \text{mod}$ $d$. Here the two systems are treated completely independently: the only ``mixing'' between them happens in the final step, since Bob's output depends on both systems' processes. The $(n_A+n_B)$-th order interference term is then
\begin{equation}
\begin{split}
\tilde{I}_{n_A+n_B}^{(d)}&=\frac{1}{d^{n_A+n_B}} \sum_{x_1,...x_{n_A+n_B}} P\left(b=s_{\vec{x}}^{(d)}|\vec{x}\right) - \frac{1}{d}
=\frac{1}{d^{n_A+n_B}} \sum_{x_1,...x_{n_A+n_B}} \sum_l P\left(b^{(A)}_l,b^{(B)}_l|\vec{x}\right) - \frac{1}{d},
\end{split}
\end{equation}
where the sum runs over all two-outcome measurements which produce the correct modulo, i.e. $\left(b^{(A)}_l+b^{(B)}_l\right) \text{mod }d=\left(\sum_{j=1} x_j\right)\text{mod}$ $d$. Intuitively, Bob can produce the correct overall modulo, even if he measures wrong single-system moduli: e.g. for $d=2$, if Bob's single-system outcomes are $s_A \oplus 1$ and $s_B \oplus 1$, where $s_A$ and $s_B$ are the correct parities, he will still produce the correct overall parity, since $(s_A \oplus 1) \oplus (s_B \oplus 1)= s_A \oplus s_B$. Indeed, for arbitrary $d$, there are in total $(d-1)$ wrong single-system outcomes which provide the correct overall output. The interference term is thus 
\begin{equation}\label{express}
\tilde{I}_{n_A+n_B}^{(d)}=q_A q_B+\sum_{i=1}^{d-1}p_i^{(A)}p_{d-i}^{(B)}- \frac{1}{d},
\end{equation} 
where $q_A$ and $q_B$ are average winning probabilities for the two systems, while the distribution averages over wrong outcomes are defined as
\begin{equation}
p_i^{(A)}\equiv \frac{1}{d^{n_A}}\sum_{x_1,...,x_{n_A}}P^{(A)}\left(b^{(A)}=(s_{\vec{x}}^{(d)}+i)\text{mod } d|\vec{x}\right),
\end{equation} 
and analogously for system B. Next, we define weights $\alpha_j$ and $\beta_j$ in the following way
\begin{equation}\label{alpha}
\begin{split}
p_i^{(A)}=(1-q_A)\alpha_i,\\
p_{d-i}^{(B)}=(1-q_B)\beta_i,
\end{split}
\end{equation}
which implies $\sum_j \alpha_j= \sum_j \beta_j =1$\footnote{In order for the coefficients $\alpha_i$ and $\beta_i$ to be well defined, we assume that $q_A\neq 1$ and $q_B\neq 1$. In the case that either $q_A$ or $q_B$ (or both) are equal to 1, the first term in \eqref{express} is $q_Aq_B>1/d$, while the second term is non-negative, which makes the overall interference term positive.}. Plugging expressions \eqref{alpha} into the interference term we obtain
\begin{equation}
\tilde{I}_{n_A+n_B}^{(d)}=q_A q_B + (1-q_A)(1-q_B)\lambda -\frac{1}{d}, \quad \lambda\equiv \sum_j \alpha_j \beta_j.
\end{equation}
By a straightforward application of the method of Lagrange multipliers with constraints $\sum_j \alpha_j=1$ and $\sum_j \beta_j =1$, one obtains that the minimum value of $\lambda$ is $\frac{1}{d-1}$ (which is achieved for uniform $\alpha_j, \beta_j$). Therefore the following inequality holds:
\begin{equation}
\tilde{I}_{n_A+n_B}^{(d)} \geq q_A q_B + \frac{1}{d-1}(1-q_A)(1-q_B) -\frac{1}{d}.
\end{equation}
The RHS of the latter inequality is equal to zero for $q_A=q_B=1/d$ and is monotonically increasing for $q_A>1/d, q_B>1/d$. Thus we constructed a process such that, if the two systems exhibit $n_A$-th and $n_B$-th order interference, the composite system exhibits $(n_A+n_B)$-th order interference.\\
Using the same reasoning, the discussion can be generalized to an arbitrary number of systems, since systems can be composed two-by-two. Therefore, by the same construction, $k$ systems of orders $\left\{n_1,...,n_k\right\}$ can produce $(\sum_{i=1}^{k} n_i)$-th order interference.

\end{document}